\documentclass[11pt]{article}

\usepackage[english]{babel}
\usepackage{amsthm}
\usepackage{algorithm}
\usepackage{algpseudocode}
\usepackage{listings}
\usepackage{hyperref} 
\usepackage{amsmath}
\usepackage{graphicx}
\usepackage{caption}
\usepackage{subcaption}
\usepackage{amssymb}
\usepackage{amsfonts}
\usepackage{bbold}
\usepackage{float}
\usepackage[margin=1.2in]{geometry}
\usepackage[square,authoryear]{natbib}
\usepackage[affil-it]{authblk}
\usepackage{fancyhdr}
\usepackage{comment}
\usepackage{xcolor}
\usepackage{multirow}
\usepackage{diagbox}
\usepackage{tikz}
\usetikzlibrary{quotes,angles}
\usepackage{pgfplots}
\usepackage{authblk}
\pgfplotsset{compat=1.18}
\usepackage{array}
\usepackage{booktabs}
\pgfplotsset{compat = newest}
\usetikzlibrary{positioning,arrows.meta}
\hypersetup{
	colorlinks=true,
	citecolor=blue  
}
\fancyhfoffset[L]{1cm} 
\fancyhfoffset[R]{1cm} 
\newcommand {\R}{\mathbb{R}}

\numberwithin{equation}{section}

\theoremstyle{definition}

\theoremstyle{remark}

\theoremstyle{remark}

\title{
	\textbf{Spatial constraints improve filtering of measurement noise from animal tracks}}

\author[1]{Alexandre Delporte\thanks{\texttt{alexandre.delporte@univ-grenoble-alpes.fr}}}
\author[2]{Susanne Ditlevsen}
\author[1]{Adeline Samson}

\affil[1]{Laboratoire Jean Kuntzmann,
	Université Grenoble-Alpes}
\affil[2]{Department of Mathematical Sciences,
	University of Copenhagen}

\date{\today}

\begin{document}
\thispagestyle{empty}
\maketitle
	
	\begin{abstract}
    
	    Advances in tracking technologies for animal movement require new statistical tools to better exploit the increasing amount of data. Animal positions are usually calculated using the GPS or Argos satellite system and include potentially  non-Gaussian and heavy-tailed measurement error patterns. Errors are usually handled through a Kalman filter algorithm, which can be sensitive to non-Gaussian error distributions.
        We introduce a latent movement model through an underdamped Langevin stochastic differential equation (SDE) that includes an additional drift term to ensure that the animal remains in a known spatial domain of interest. This can be applied to aquatic animals moving in water or terrestrial animals moving in a restricted zone delimited by fences or natural barriers. We demonstrate that the incorporation of these spatial constraints into the latent movement model can improve the accuracy of filtering for noisy observations of the positions. We implement an Extended Kalman Filter as well as a particle filter adapted to non-Gaussian error distributions. Our filters are based on solving the SDE through splitting schemes to approximate the latent dynamic. We illustrate the approach on a real Argos telemetry track of a bowhead whale in Foxe Basin, Canada.
        
	\end{abstract}
    
	\section{Introduction}

Animal tracking allows us to uncover movement patterns of wild species in their natural environment. Having accurate trajectories is of particular interest to better infer space use and define conservation areas \citep{hays_high_2021}. 
Position data typically comes with measurement error whose distribution depends on the type of device attached to the animal and the method used to estimate the position. 
One of the most common methods for tracking animals at sea is via the Argos telemetry system, with more than $40 000$ individuals tracked since 2007 \citep{jonsen_continuous-time_2020}. In this system, the estimation of the animal position, which relies on Doppler shift calculations, can exhibit heavy-tailed and anisotropic errors of several kilometers \citep{hoenner_enhancing_2012}. Alternatively, new Fastloc-GPS loggers deployed in wild animals have provided more accurate positions with heavy-tailed errors ranging from ten to a few hundreds meters \citep{wensveen_path_2015}. 


To estimate true positions from noisy observations, it is essential to accurately model the error distribution, as well as the latent dynamics of motion.
In the literature, the estimation of the true position given the observations is typically performed using a Kalman filter algorithm \citep{johnson_continuous-time_2008,jonsen_continuous-time_2020,michelot_varying-coefficient_2021}. This algorithm may be sensitive to non-Gaussian errors and requires to pre-filter the data to ensure that it does not deviate too much from Gaussianity \citep{patterson_using_2010}.
This motivates the development of statistical methodology that can handle flexible measurement error distributions, associated with a realistic latent movement model.

Stochastic differential equations (SDEs) have gained increasing attention in movement ecology for modelling animal trajectories in continuous time and space \citep{johnson_continuous-time_2008}. A popular class of models is based on the overdamped Langevin equation, where the drift is the gradient of a potential function that represent attractive areas. \cite{gloaguen_stochastic_2018} define such a model with a multimodal Gaussian mixture potential and apply it to infer fishing zones from vessel trajectories. \cite{michelot_langevin_2019} employ a similar overdamped Langevin SDE to analyze the movement of Steller sea lions in Alaska. However, these frameworks do not account for measurement errors in the observations. An important additional feature of animal movement is that trajectories are constrained by landscape boundaries that animals cannot cross. \cite{cholaquidis_level_2020} use a reflected Brownian motion with drift to estimate home ranges and core areas of elephants, but do not consider measurement error. A common approach to incorporate spatial constraints  consists in specifying a potential function with a steep gradient near the boundary of the domain, so that the drift pushes the animal back toward the interior when it approaches the edge \citep{brillinger_simulating_2003,hooten_animal_2017}. In practice, this specification is straightforward only for simple domains such as disks or rectangles \citep{russell_spatially_2018,hooten_animal_2017}, and we are not aware of applications to more complex domains. Fewer works combine spatial constraints with measurement error. \cite{hanks_reflected_2017} define a reflected linear SDE and infer movement parameters from Argos data of sea lion trajectories with non-Gaussian location errors, incorporating spatial constraints by projecting positions that fall outside the domain onto its boundary. \cite{delporte_varying_2025} model spatial constraints through a smooth rotation of the velocity as positions approach the domain's boundary, assuming Gaussian measurement errors and inferring movement parameters through a classical Kalman filter. To our knowledge, no existing work combines a nonlinear SDE latent model with spatial constraints and non-Gaussian measurement errors.

We propose a framework that addresses these three aspects simultaneously. The central contribution is a penalized overdamped Langevin SDE that enforces spatial constraints on polygon domains. The penalization term is identically zero when the position lies inside the domain or on its boundary, and only orients the velocity inward when the animal has stepped outside. The underlying Langevin dynamics are therefore unaffected by the boundary constraint during normal movement. For a convex domain, the penalty term is equivalent to a repulsive potential proportional to the squared distance to the closure of the domain, making the connection to the potential-based approach of \cite{hooten_animal_2017} explicit. This approach has direct theoretical grounding as a penalized approximation to a reflected SDE \citep{liu_discretization_1995}, which provides a basis for the calibration of the penalty parameter $\lambda$
\citep{pettersson_penalization_1997}. Beyond the boundary term, we specify a nonlinear drift through a multimodal potential defined as a mixture of Gaussians, in a manner similar to \cite{gloaguen_stochastic_2018}, to represent preferred habitat zones, and solve the resulting nonlinear SDE using splitting schemes adapted to that nonlinearity. We implement both an extended Kalman filter for Gaussian measurement errors and a particle filter to handle non-Gaussian heavy-tailed errors such as Student's
t-distributions.

Through a simulation study, we compare the performances of Kalman filtering, extended Kalman filtering, and particle filtering, both with and without spatial penalization. We demonstrate that incorporating domain knowledge can improve filtering accuracy, especially when the observation frequency is high or the measurement error distribution deviates from Gaussianity. This is confirmed on real Argos telemetry data of a bowhead whale.

The remainder of the paper is structured as follows. In Section~\ref{sec: penalized Langevin}, we introduce the penalized Langevin SDE. In Section~\ref{sec: measurement error model}, we describe the measurement error models that we consider throughout the paper. Section~\ref{sec: splitting schemes} presents numerical splitting schemes for simulating the dynamics. Section~\ref{sec: filtering algorithms} details the filtering algorithms under different measurement error specifications. It includes Kalman filtering and extended Kalman filtering for Gaussian measurement error, as well as particle filtering for Student's t-distributed errors. Section~\ref{sec: simulation study} is dedicated to a simulation study in different scenarios. We vary the measurement error distribution as well as the time step between consecutive observations and compare the performances of the filtering algorithms. Then, we provide an example of application to bowhead whale telemetry data in Section~\ref{sec:bowhead} and compare the particle filter and the Kalman filter results. We conclude with a discussion of potential applications and extensions.

\section{Penalized Langevin SDE in a bounded domain}
\label{sec: penalized Langevin}

\subsection{Stochastic differential equation}
Let $\mathcal{D} \subset \R^2$ be an open bounded domain representing for instance water areas, or a known home range of the species whose movement we are interested in. We denote by $(X(t))_{t \geq 0}$ the position process and $(V(t))_{t \geq 0}$ the velocity process, both in two dimensions. We suppose they solve the following penalized Langevin SDE:
\begin{equation}
	\begin{cases}
		dX(t)=V(t) dt \\
		dV(t)=-AV(t)dt-\nabla H(X(t))dt+\sigma dW(t)-\beta_\lambda(X(t)) dt
	\end{cases}
\label{eq: penalized Langevin}
\end{equation}
where $A$ is a positive definite matrix parametrised as
\[A=\begin{pmatrix} c & -\omega \\ \omega & c\end{pmatrix}\]
with $c>0$ a damping parameter and $\omega$ an angular velocity parameter. 
 $W(t), t \geq 0$, is a two-dimensional Brownian motion. 
 In movement ecology, a Langevin SDE such as~\eqref{eq: penalized Langevin} are often parametrised in terms of the persistence time $\tau=\frac{1}{c}$ and velocity $\nu = \frac{\sigma}{2} \sqrt{\pi \tau}$ \citep{gurarie_correlated_2017}. The persistence captures the autocorrelation of the velocity, and $\nu$ controls the mean velocity norm. The parameter $\omega$ represents a rotational tendency in the movement, which is supposed to be constant here.

\subsection{Penalization}

The penalization term $\beta_\lambda(X(t))$ is given by 
\[\beta_\lambda(x)=\frac{x-\pi(x)}{\lambda} \quad \text{and } \pi(x)=\underset{x' \in \overline{\mathcal{D}}}{\mathrm{argmin} } \vert \vert x-x'\vert \vert, \quad x \in \R^2,\]
where $\pi (x)$ is the projection of the position $x$ on $\overline{\mathcal{D}}$, the closure of $\mathcal{D}$. This means that when $x$ belongs to $\mathcal{D}$ or its boundary, $\beta_{\lambda}=0$ and the penalization term does not influence the velocity. However, when $x$ is strictly outside $\mathcal{D}$ and its boundary, the penalization becomes non zero and points outward, so that the term $-\beta_{\lambda}(x)$ reorients the velocity inward and the position process $X$ smoothly re-enters the domain $\mathcal{D}$. 
The parameter $\lambda$ controls the strength of the penalization applied to keep the process inside $\mathcal{D}$. A small value of $\lambda$ induces a large drift $\beta_\lambda$, immediately and strongly pushing the position inward as soon as it leaves the domain, while a larger value of $\lambda$ induces a smaller drift and hence a smaller correction, where the animal might spend more time outside the domain.  When the domain $\mathcal{D}$ is convex, the penalization simplifies to \citep{liu_discretization_1995}
$$\beta_\lambda(x)=\frac{1}{2\lambda} \nabla_x \underset{x' \in \overline{\mathcal{D}}}{\inf} \vert \vert x-x'\vert \vert^2.$$
Figure ~\ref{fig: penalization vector field} illustrates the penalization for a disk. In this case, the penalization could be written in closed form. However, in practice $\mathcal{D}$ can represent an area delimited by fences that the animal cannot cross \citep{brillinger_simulating_2003}, as well as water areas surrounded by coastlines that restrict the movement of marine mammals \citep{hanks_reflected_2017}, which are better represented by a polygon or multiple polygons than a simple disk. 
If we suppose that $\mathcal{D}$ is a polygon with vertices $(v_k)_{1 \leq k \leq p}$, then we can compute the projection $\pi(x)$ similarly to \citep{hanks_reflected_2017}.
For $x \notin \overline{\mathcal{D}}$, there exists a vertex $v_k$ such that $\pi(x)=v_k+\gamma(x)(v_{k+1}-v_k)$ where $\gamma(x)=\max(0,\min(1,\frac{\langle x-v_k,v_{k+1}-v_k\rangle}{l^2}))$ and $l=\vert\vert v_{k+1}-v_k\vert \vert$. In theory, the probability that $\pi(x) \in \{v_k,v_{k+1}\}$ is zero, in which case we have $\gamma(x)=\frac{\langle x-v_k,v_{k+1}-v_k\rangle}{l^2}$. Hence, the projection operator $\pi$ is piecewise linear in $x$. 

	\begin{figure}[ht!]
		\centering
		\begin{subfigure}[t]{0.45\textwidth}
			\centering
			\begin{tikzpicture}
				\begin{axis}[
					xmin = -4, xmax = 4,
					ymin = -4, ymax = 4,
					axis equal image,
					view = {0}{90},
					xlabel = {$x_1$},
					ylabel = {$x_2$},
					legend style={at={(1,1)},anchor=north east},
					]
					
					\addplot[
					fill=blue!30,
					draw=none,
					domain = 0:360,
					samples = 100
					] ({cos(x)}, {sin(x)});
					
					\addplot3[
					quiver = {
						u = {(1 - 1/sqrt(x^2 + y^2)) * x},
						v = {(1 - 1/sqrt(x^2 + y^2)) * y},
						scale arrows = 0.15,
					},
					stealth-,
					samples = 17,
					domain = -4:4,
					domain y = -4:4,
					restrict expr to domain={sqrt(x^2 + y^2)}{1:4},
					] {0};
					
					\addlegendentry{Vector field $-\beta_1(x_1, x_2)$}
					
					\addplot[
					thick,
					black,
					domain = 0:360,
					samples = 100
					] ({cos(x)}, {sin(x)});
					
				\end{axis}
			\end{tikzpicture}
			\caption{}
		\end{subfigure}
		\hfill
		\begin{subfigure}[t]{0.45\textwidth}
			\centering
			\begin{tikzpicture}
				\begin{axis}[
					xmin = -4, xmax = 4,
					ymin = -4, ymax = 4,
					axis equal image,
					view = {0}{90},
					xlabel = {$x_1$},
					ylabel = {$x_2$},
					legend style={at={(1,1)},anchor=north east},
					]
					\addplot[
					fill=blue!30,
					draw=none,
					domain = 0:360,
					samples = 100
					] ({cos(x)}, {sin(x)});
					
					\addplot3[
					quiver = {
						u = {(1 - 1/sqrt(x^2 + y^2)) * x*10},
						v = {(1 - 1/sqrt(x^2 + y^2)) * y*10},
						scale arrows = 0.15
					},
					stealth-,
					samples = 17,
					domain = -4:4,
					domain y = -4:4,
					restrict expr to domain={sqrt(x^2 + y^2)}{1:4},
					] {0};
					
					\addlegendentry{Vector field $-\beta_{0.1}(x_1, x_2)$}
					
					\addplot[
					thick,
					black,
					domain = 0:360,
					samples = 100
					] ({cos(x)}, {sin(x)});
					
				\end{axis}
			\end{tikzpicture}
			\caption{}
		\end{subfigure}
		\caption{Vector fields $-\beta_\lambda$ for (a) $\lambda=1$ and (b) $\lambda=0.1$. Lower values of $\lambda$ produce a stronger inward push.}
        \label{fig: penalization vector field}
	\end{figure}

\subsection{Potential surface}
The function $H$ defines the potential surface that controls areas of attraction (or repulsion) within the domain $\mathcal{D}$. Examples of potential surfaces $H$ can be found in \cite{preisler_modeling_2004}. In the sequel, we consider a potential surface $H$ in~\eqref{eq: penalized Langevin} defined as a mixture of Gaussian potentials similar to \cite{gloaguen_stochastic_2018}:
\[\forall x \in \R^2 \, : \quad H(x)=-\sum_{j=1}^J H_j(x) \mbox{ with } H_j(x)= \alpha_j \exp(-(x-x^*_j)^\top B_j(x-x^*_j))\]
with weights $\alpha_j$, precision matrices $B_j \in \mathcal{S}_2^+$, the set of $2\times 2$ symmetric positive definite matrices,  and centers $x^*_j \in \mathcal{D}$ for $j \in \{1,\ldots,J\}$. For weights $\alpha_j >0$, the negative sign in the definition of $H$ ensures that the drift term $-\nabla H(X(t))$ points toward the centers $x^*_j$ , so each Gaussian component acts as an attractive potential well. Ecologically, the centers $x^*_j$ may represent preferred areas such as foraging or resting sites, the weights $\alpha_j$ control the strength of attraction toward each site, and the precision matrices $B_j$ control the spatial extent of attraction in each direction.
We denote $e_j(x)=\exp(-(x-x^*_j)^\top B_j(x-x^*_j))$. 
The gradient and Hessian matrices are given by (see Appendix \ref{App:A}):
\begin{equation}\nabla H(x)=\sum_{j=1}^J 2\alpha_j B_j (x-x^*_j) e_j(x)
	\label{eq: potential gradient}
\end{equation}

\begin{equation}
	D^2H(x)=\sum_{j=1}^J 2\alpha_je_j(x)B_j\left(I_2-2(x-x^*_j)(x-x^*_j)^\top B_j\right).
	\label{eq: potential hessian}
\end{equation}
The model proposed by \cite{gurarie_correlated_2017} or \cite{johnson_continuous-time_2008} corresponds to a flat  surface $H$ with no penalization ($\lambda =+\infty$).

\section{Measurement error models}
\label{sec: measurement error model}
In this section, we describe the statistical models used to represent measurement errors in the observed animal positions. Accurate modeling of these errors is essential for reliable state estimation, as their distribution strongly depends on the tracking technology and can affect the performance of filtering algorithms.

In practice, the velocity process $V$ is not observed, only the position process $X$ is observed with measurement error.
We denote $(y_j)_{1 \leq j \leq n}$ the discrete-time noisy observations of the positions. 

\subsection{Gaussian measurement errors}
The most common assumption for animal tracks is Gaussian error distributions \citep{johnson_continuous-time_2008}. The reason it is so popular is that under this assumption, filtering as well as parameters estimation are easily done with the Kalman filter \citep{michelot_varying-coefficient_2021}.  The Gaussian measurement error model is 
\begin{equation}
		y_j=LU_j+\varepsilon_j, \quad \varepsilon_j \sim \mathcal{N}(0,\sigma_{obs}^2I_2),
\end{equation}
where $L=\begin{pmatrix} I_2 & 0_2\end{pmatrix}$ and $(U_j)_{1 \leq j \leq n}=\begin{pmatrix} X_j & V_j \end{pmatrix}_{1 \leq j \leq n}$ is the vector of true positions and velocities at observation times. Here $I_2$ denotes the 2-dimensional identity matrix and $0_{2}$ denotes a $2 \times 2$ null-matrix. The parameter $\sigma_{obs}$ quantifies the scale of the errors in the directions Northing and Easting (isotropic).

\subsection{Isotropic Student's t-distributed measurement errors}
However, GPS tracks often exhibit heavy-tailed error distributions \citep{wensveen_path_2015}, which are best modelled with Student's t-distribution, $t_d$, where $d$ is the degrees of freedom. The measurement error model is then

\begin{equation}
	y_j=LU_j+\sigma_{obs}\varepsilon_j, \quad \varepsilon_j \sim t_d,
    \label{eq: isotropic student error}
\end{equation}
where $\sigma_{obs}$ again quantifies the isotropic error scale in Northing and Easting. In practice, both $\sigma_{obs}$ and $d$ may vary with the number of satellites that capture the signal \citep{wensveen_path_2015}. An increased number of satellites results in more accurate positions with a distribution closer to a Gaussian (i.e., larger $d$). For simplicity, we consider a constant value of $\sigma_{obs}$ and $d$ in this paragraph, however, it can easily be generalized.

\subsection{Argos X-shaped measurement errors}
For tracks relying on the Argos system, Doppler shift is used to estimate animal positions and the errors have generally higher variance and a complex shape favouring errors in the North-West/South-East and North-East/South-West directions \citep{brost_animal_2015,hanks_reflected_2017}.
This can be modeled as follows:
\begin{equation}
	y_j=LU_j+\varepsilon_j 
\end{equation}
where \[\varepsilon_j \sim \begin{cases} t_d(0,\Sigma) \mbox{ with probability } p, \\ t_d(0,\tilde{\Sigma}) \mbox{ with probability } 1-p, \end{cases}\]
with $\Sigma=\sigma_{obs}^2\begin{pmatrix} 1 & \rho\sqrt{a} \\ \rho\sqrt{a} & 1\end{pmatrix}$ and $\tilde{\Sigma}=\sigma_{obs}^2\begin{pmatrix} 1 & -\rho\sqrt{a} \\ -\rho\sqrt{a} & 1\end{pmatrix}$.
The notation $t_d(0,\Sigma)$ denotes a bivariate Student's t-distribution with $d$ degrees of freedom, mean vector $0$ and scale matrix $\Sigma$. The parameter $p$ assigns weights to the North-West/South-East and North-East/South-West directions, $\rho \in (-1,1)$ is a correlation parameter and $a >0$ is an anisotropy parameter. When $a=0$, we obtain an isotropic Student distribution similar to~\eqref{eq: isotropic student error}. Only the product $\rho \sqrt{a}$ can be identified, not each parameter.

\begin{figure}[ht!]
	\centering
	\includegraphics[scale=0.5]{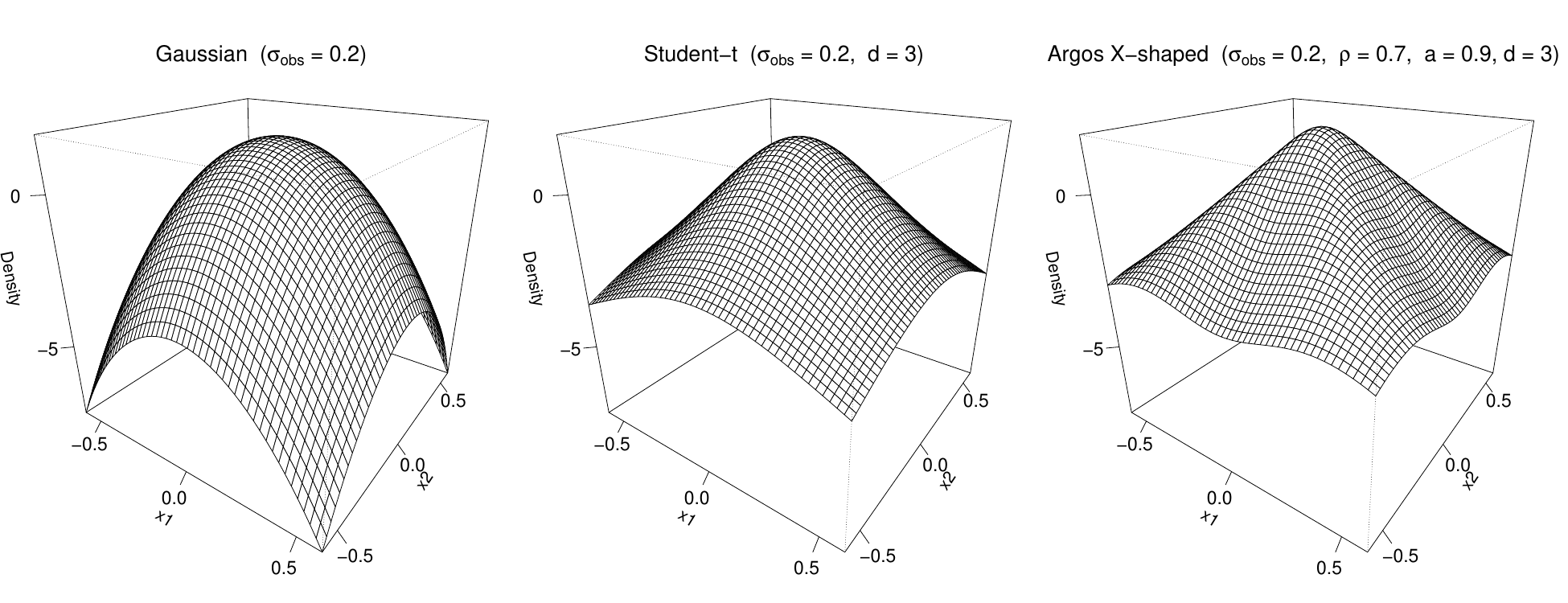}
	\vspace{-0.5cm}
	\caption{Log-density of each error distribution : Gaussian (left), Student's t-distribution (middle) and Argos X-shaped distribution with $p=0.5$ (right). All distributions have scale $\sigma_{obs}=0.2$.
	}
	\label{fig:error log density}
\end{figure}

Figure~\ref{fig:error log density} shows the log-density for each type of error distribution.
In the Gaussian case, the observed positions are distributed circularly around the true position, reflecting small and isotropic errors.
For the case of Student's-t, the majority of the observations remain close to the true path, but occasional outliers appear far from the animal's actual position due to the heavy-tailed nature of the distribution.
In the Argos case, the errors are both large and directional: observations spread preferentially along the North-West/South-East and North-East/South-West axes, producing stretched error patterns that differ strongly from the other two models.

\section{Splitting schemes}
\label{sec: splitting schemes}
With no penalization and a flat potential surface, the SDE is linear with an expact Gaussian solution. However, both the penalization term and the potential function 
$H$ introduce non-linearity into the equation and the solution is not exact. We need a numerical scheme to approximate the solution. Classical schemes such as Euler-Maruyama may not be precise enough to capture the non-linearity. We instead employ a splitting method to approximate the solution of the penalized SDE, which is particularly suitable for non-linear SDEs. The idea is to split the process into a linear SDE, for which we know the distribution, and an ordinary differential equation (ODE), which can be exactly solved \citep{pilipovic_parameter_2024}.

\subsection{SDE splitting}

Splitting methods were originally designed to solve ordinary and partial
differential equations \citep{mclachlan_splitting_2002} by decomposing a system into subsystems that are easier to solve, and combining the solutions.
For an SDE such as~\eqref{eq: penalized Langevin} with constant diffusion,
one can split the equation into an SDE with linear drift and constant
diffusion, and an ordinary differential equation with the remaining
nonlinear drift term \citep{pilipovic_strang_2025}.
In practice, it is recommended to linearize the drift in the SDE around
one of its fixed points \citep{pilipovic_parameter_2024}.
We may choose a fixed point
$u_l^* = \begin{pmatrix} x_l^{*\top} & 0_{2,1}^\top \end{pmatrix}^\top$,
$l \in \{1, \ldots, J\}$, where $x_l^*$ is one of the centers of attraction
of the potential $H$ and the velocity component is set to zero.
Hence, we rewrite the SDE as (see Appendix \ref{App:B})
\begin{equation}\label{eq: penalized Langevin rewritten}
\begin{split}
d\begin{pmatrix} X(t) \\ V(t) \end{pmatrix}
&=
\begin{pmatrix} 0_2 & I_2 \\ -2\alpha_l B_l & -A \end{pmatrix}
\Bigg[ \begin{pmatrix} X(t) \\ V(t) \end{pmatrix}
- \begin{pmatrix} x_l^* \\ 0_{2,1} \end{pmatrix} \Bigg] dt
+ \begin{pmatrix} 0_2 \\ \sigma I_2 \end{pmatrix} dW(t) \\
&\quad
+ \begin{pmatrix} 0_{2,1} \\
-2\alpha_l (e_l(X(t))-1) B_l (X(t)-x_l^*)
- \nabla H_{-l}(X(t)) - \beta_\lambda(X(t))
\end{pmatrix} dt
\end{split}
\end{equation}
where $H_{-l}(x)=-\sum_{j=1, j \neq l}^J H_j(x)$ and $0_{2,1}$ is the null column vector of size $2$. We then define the splitting through the two processes:

\begin{align}
\label{eq:XVsplit}
	\begin{cases}
		d\begin{pmatrix} X^{(1)}(t) \\ V^{(1)}(t) \end{pmatrix}=\begin{pmatrix} 0_2 & I_2 \\ -2\alpha_l B_l & -A \end{pmatrix} \left[ \begin{pmatrix} X^{(1)}(t) \\ V^{(1)}(t) \end{pmatrix} - \begin{pmatrix}  x_l^* \\ 0_{2,1}\end{pmatrix}\right] dt+\begin{pmatrix} 0_{2} \\ \sigma I_2 \end{pmatrix} dW(t)
		\\
		d\begin{pmatrix} X^{(2)}(t) \\ V^{(2)}(t) \end{pmatrix}=\begin{pmatrix} 0_{2,1} \\ -2\alpha_l (e_l(X^{(2)}(t))-1) B_l (X^{(2)}(t)-x_l^*)-\nabla H_{-l}(X^{(2)}(t))-\beta_\lambda(X^{(2)}(t))\end{pmatrix}dt
	\end{cases}
\end{align}
The splitting approximation combines the solutions of the two equations.
 Denote the solutions of the SDE and ODE with initial condition $u=\begin{pmatrix} x & v \end{pmatrix}^\top \in \R^4$ by $\phi_h^{(1)}(u)$ and $\phi_h^{(2)}(u)$, respectively, over a time-interval of length $h$. 

In the following, we detail the solutions $\phi_h^{(1)}$ and $\phi_h^{(2)}$ of the two systems.

\subsection{Solutions of the two systems}

The first process in \eqref{eq:XVsplit} is a four-dimensional Ornstein-Uhlenbeck process with exact solution
\[\begin{pmatrix}X^{(1)}(t+h) \\ V^{(1)}(t+h) \end{pmatrix} =e^{\tilde{A}h}\left(\begin{pmatrix}X^{(1)}(t) \\ V^{(1)}(t) \end{pmatrix}- \begin{pmatrix}  x_l^* \\ 0_{2,1}\end{pmatrix}\right) + \begin{pmatrix}  x_l^* \\ 0_{2,1}\end{pmatrix}+ \eta(h)\]
with $\tilde{A}=\begin{pmatrix} 0_2 & I_2 \\ -2\alpha_l B_l & -A\end{pmatrix}$,  $\eta(h) \sim \mathcal{N}(0_{4,1},\tilde{Q}(h))$ and 
$$\tilde{Q}(h)=\int_0^h e^{-\tilde{A}(u-h)} \Gamma   e^{-\tilde{A}^\top(u-h)} du \, ; \quad\Gamma= \begin{pmatrix} 0_2 & 0_2 \\ 0_2 & \sigma^2 I_2\end{pmatrix}.$$
The covariance matrix $\tilde{Q}(h)$ is defined as an integral which can be computed exactly. 
Writing $C$ for the matrix such that $\mathrm{vec}(C)=(\tilde{A}\oplus \tilde{A})^{-1} \mathrm{vec}(\Gamma)$, where $\oplus$ is the Kronecker sum, and $\mathrm{vec}$ is the vectorization operator, we have (see Appendix \ref{App:C})
\begin{equation}
\tilde{Q}(h)=e^{\tilde{A} h} Ce^{\tilde{A}^\top h}-C.
\label{eq: OU cov matrix}
\end{equation}
The second process in \eqref{eq:XVsplit} is an ODE with exact solution $\phi^{(2)}_h(u)=u-hg_\lambda(u)$ where $$g_\lambda(u)=g_\lambda(x,v)=\begin{pmatrix}0_{2,1} \\ 2\alpha_l(e_l(x)-1)B_l(x-x^*_l)+\nabla H_{-l}(x)+\beta_\lambda(x) \end{pmatrix}.$$ Note that the function $g_\lambda$ depends only on $x$ and not on $v$.

 There are several ways to combine these solutions that each gives different approximations. We present here only the so-called Lie-Trotter and Strang approximations.
 
\subsection{Lie-Trotter and Strang approximations}
Denote by $U=\begin{pmatrix} X^\top & V^\top \end{pmatrix} \in \R^4$ the joint process of positions and velocities solving~\eqref{eq: penalized Langevin}.  
 The Lie-Trotter splitting provides the approximation $U^{[\mathrm{LT}]}$ iteratively defined by: 
\[U^{[\mathrm{LT}]}(t+h)=\phi_h^{(1)} \circ \phi_h^{(2)}(U^{[\mathrm{LT}]}(t)).\]

At time $t+h$, given the current value $U^{[\mathrm{LT}]}(t)$ at time $t$ , the next position and velocity are
\[U^{[\mathrm{LT}]}(t+h)=e^{\tilde{A}h}\left(U^{[\mathrm{LT}]}(t)-u_l^*-hg_{\lambda}(U^{[\mathrm{LT}]}(t))\right)+u_l^*+\eta(h)\]
where $\eta(h) \sim \mathcal{N}(0_{4,1},\tilde{Q}(h))$. Hence, $U^{[\mathrm{LT}]}(t+h)$ is Gaussian conditionally on $U^{[\mathrm{LT}]}(t)$.

The Strang splitting provides the approximation $U^{[\mathrm{S}]}$ defined iteratively by: 
\[U^{[\mathrm{S}]}(t+h)=\phi_{\frac{h}{2}}^{(2)} \circ \phi_h^{(1)} \circ \phi_{\frac{h}{2}}^{(2)}(U^{[\mathrm{S}]}(t)).\]
At time $t+h$, the next position and velocity given the current value $U^{[\mathrm{S}]}(t)$ are
\begin{equation}U^{[\mathrm{S}]}(t+h)=Z-\frac{h}{2}g_{\lambda}(Z), \quad  Z=e^{\tilde{A}h} (\hat{U}_{\frac{h}{2}}-u_l^*)+u_l^*+\eta\left(h\right),  \quad \hat{U}_{\frac{h}{2}}=U^{[\mathrm{S}]}(t)-\frac{h}{2}g_{\lambda}(U^{[\mathrm{S}]}(t)).
\label{eq: Strang approximation}
\end{equation}
Contrary to the Lie-Trotter approximation, $U^{[\mathrm{S}]}(t+h)$ is not Gaussian conditionally on $U^{[\mathrm{S}]}(t)$, only $Z:=\begin{pmatrix}X_Z& V_Z \end{pmatrix}^\top$ is conditionally Gaussian with mean $\tilde{m}(h)=e^{\tilde{A}h} (\hat{U}_{\frac{h}{2}}-u_l^*)+u_l^*$ and covariance matrix $\tilde{Q}(h)$ \citep{pilipovic_parameter_2024}.
Since the first two coordinates of $g_{\lambda}$ are zero, we have that $X^{[\mathrm{S}]}(t+h)=X_Z$, implying that the marginal distribution of $X^{[\mathrm{S}]}(t+h)$ conditionally on $U^{[\mathrm{S}]}(t)$ is Gaussian with
$\tilde{m}_x$ and covariance $\tilde{Q}_{xx}$, where we denote $\tilde{m}(h)=\begin{pmatrix} \tilde{m}_x(h)^\top & \tilde{m}_v(h)^\top \end{pmatrix}\top$ and $ \tilde{Q}(h)=\begin{pmatrix} \tilde{Q}_{xx} & \tilde{Q}_{xv} \\ \tilde{Q}_{vx} & \tilde{Q}_{vv}\end{pmatrix}$ the mean and covariance matrix.
Moreover, by~\eqref{eq: Strang approximation},
\begin{equation*}
    V^{[\mathrm{S}]}(t+h)=V_Z-\frac{h}{2} 2\alpha_l(e_l(X_Z)-1)B_l(X_Z-x^*_l)+\nabla H_{-l}(X_Z)+\beta_\lambda(X_Z).
\end{equation*}
Therefore, conditionally on $U^{[\mathrm{S}]}(t)$ and $X^{[\mathrm{S}]}(t+h)=x$  , $V^{[\mathrm{S}]}(t+h)$ is also Gaussian with mean $\tilde{m}_{v \vert x}$ and covariance matrix $\tilde{Q}_{v \vert x}$ given by
\[\tilde{m}_{v  \vert x}=\tilde{m}_v+\tilde{Q}_{vx}\tilde{Q}_{xx}^{-1}(x-\tilde{m}_x)-\frac{h}{2}(2\alpha_l(e_l(x)-1)B_l(x-x^*_l)+\nabla_{-l} H(x)) +\beta_\lambda(x), \]
\[\tilde{Q}_{v \vert x}=\tilde{Q}_{vv}-\tilde{Q}_{vx} \tilde{Q}_{xx}^{-1} \tilde{Q}_{xv}.\] 
These numerical approximations will be at the core of the filtering algorithms, and they also provide a direct way to simulate paths of the process $U$. 
\subsection{Simulation of an approximate path}
\label{subsec: approximate path sim}

In practice, we choose the fixed point $x^*_l$ in the splitting scheme depending on the current location $X(t)$.
We choose the center $l$ that has the steepest gradient at the current location, that is $l=\underset{j \in \{1,\cdots, J\}}{\mathrm{argmax}}  \log(\vert \vert \nabla H_j(X(t)) \vert \vert^2)$ where
$$ \log(\vert \vert \nabla H_j(X(t)) \vert \vert^2)=2\log(2)+2\log(\alpha_j)-2(x-x^*_j)^\top B_j(x-x^*_j)+\log((x-x^*_j)^\top B_j^\top B_j(x-x^*_j)).$$

Given the current state $U(t) = \begin{pmatrix} X(t)^\top & V(t)^\top
\end{pmatrix}^\top$, both the Lie-Trotter and Strang schemes proceed by
first applying the ODE step $\phi_h^{(2)}$ (Lie-Trotter) or a half ODE
step $\phi_{h/2}^{(2)}$ (Strang), then the linear SDE step $\phi_h^{(1)}$,
which requires sampling a Gaussian increment $\eta(h) \sim
\mathcal{N}(0, \tilde{Q}(h))$. The Strang scheme then applies a final half ODE step.
Algorithms~\ref{alg:LT} and~\ref{alg:S} summarize the two procedures. The matrix exponential in the SDE steps is computed numerically via the \texttt{expm} R package \citep{maechler_expm_2010}, and the covariance matrix $\tilde{Q}$ is computed via~\eqref{eq: OU cov matrix}.
 
\begin{algorithm}[ht!]
\caption{Lie-Trotter splitting --- one step from $U^{[\mathrm{LT}]}(t)$ to $U^{[\mathrm{LT}]}(t+h)$}
\label{alg:LT}
\begin{algorithmic}[1]
    \Require Current state $U^{[\mathrm{LT}]}(t)$, step size $h$, fixed points
             $\{x_l^*\}_{l=1}^J$, matrix $\tilde{A}$
    \State Select active center:
           $l \leftarrow \underset{j}{\mathrm{argmax}}\,
           \log\|\nabla H_j(X^{[\mathrm{LT}]}(t))\|^2$
    \State ODE step:
           $\hat{U}_ h \leftarrow U^{[\mathrm{LT}]}(t) - h\, g_\lambda(U^{[\mathrm{LT}]}(t))$
    \State Sample Gaussian increment:
           $\eta \sim \mathcal{N}(0_{4,1},\, \tilde{Q}(h))$
    \State SDE step:
           $U^{[\mathrm{LT}]}(t+h) \leftarrow
           e^{\tilde{A}h}\!\left(\hat{U}_ h - u_l^*\right) + u_l^* + \eta$
    \State \Return $U^{[\mathrm{LT}]}(t+h)$
\end{algorithmic}
\end{algorithm}
 
\begin{algorithm}[ht!]
\caption{Strang splitting --- one step from $U^{[\mathrm{S}]}(t)$ to $U^{[\mathrm{S}]}(t+h)$}
\label{alg:S}
\begin{algorithmic}[1]
    \Require Current state $U^{[\mathrm{S}]}(t)$, step size $h$, fixed points
             $\{x_l^*\}_{l=1}^J$, matrix $\tilde{A}$
    \State Select active center:
           $l \leftarrow \underset{j}{\mathrm{argmax}}\,
           \log\|\nabla H_j(X(t)^{[\mathrm{S}]})\|^2$
    \State First half ODE step:
           $\hat{U}_{h/2} \leftarrow U^{[\mathrm{S}]}(t) - \tfrac{h}{2}\, g_\lambda(U^{[\mathrm{S}]}(t))$
    \State Sample Gaussian increment:
           $\eta \sim \mathcal{N}(0_{4,1},\, \tilde{Q}(h))$
    \State SDE step:
           $Z \leftarrow
           e^{\tilde{A}h}\!\left(\hat{U}_{h/2} - u_l^*\right) + u_l^* + \eta$
    \State Second half ODE step:
           $U^{[\mathrm{S}]}(t+h) \leftarrow Z - \tfrac{h}{2}\, g_\lambda(Z)$
   \State \Return $U^{[\mathrm{S}]}(t+h)$
\end{algorithmic}
\end{algorithm}
 
\begin{figure}[ht!]
	\centering
	\includegraphics[scale=0.5]{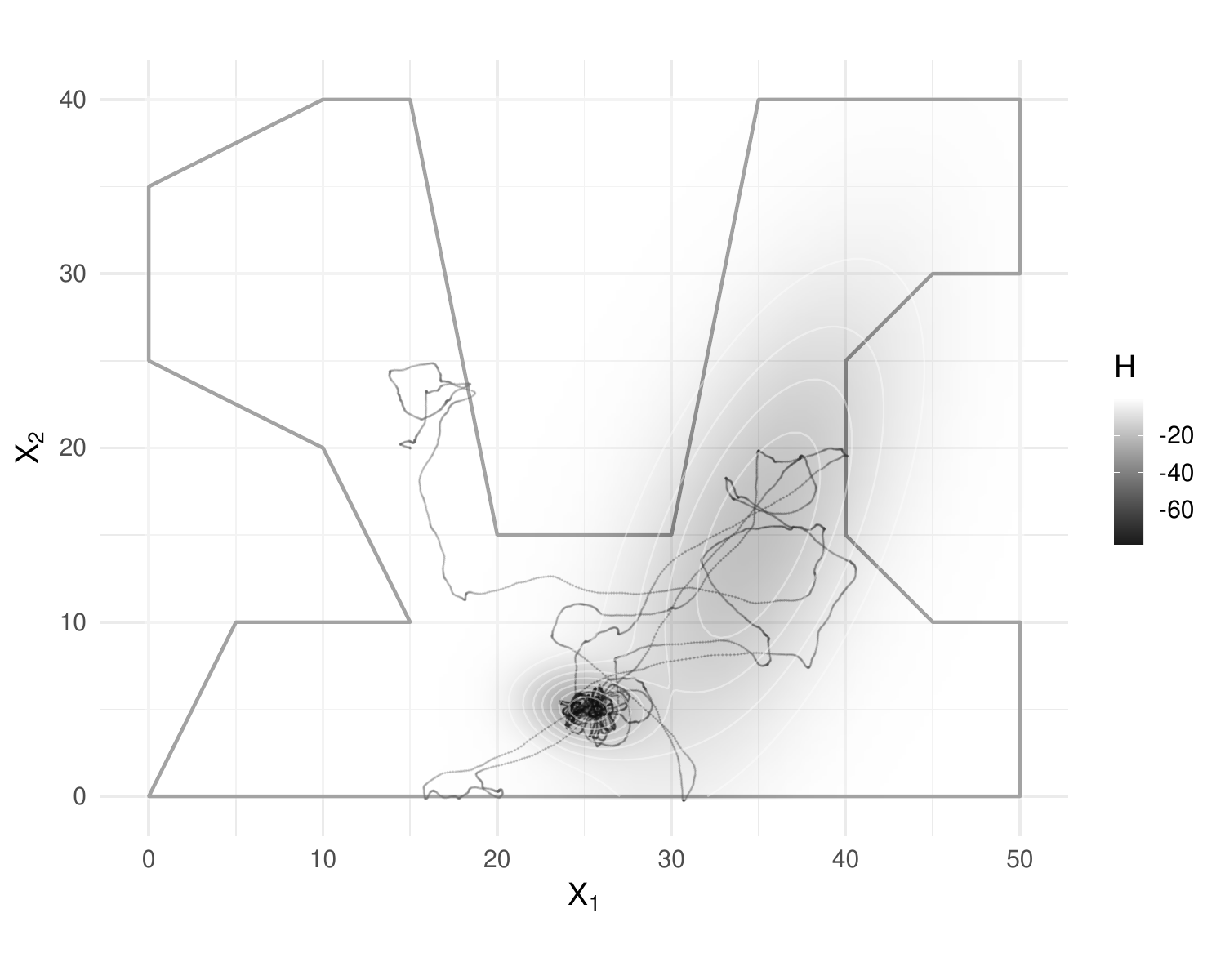}
	\vspace{-0.5cm}
	\caption{Trajectory from  \eqref{eq: penalized Langevin} with a mixture of Gaussian potentials using the Lie-Trotter approximation scheme. Parameters are $h=1/3600$, $\lambda=h^{0.8}$, $\tau=1$, $\nu=5$, $\omega=0.1$.}
	\label{fig:trajectory mix gaussian potential}
\end{figure}

Figure~\ref{fig:trajectory mix gaussian potential} shows a trajectory of \eqref{eq: penalized Langevin} with a mixture of two Gaussian potentials around $x^*_1=(25,5)$ and $x^*_2=(35,15)$ obtained with the Lie-Trotter approximation with time step $h=1/3600$ hour ($1$ second between consecutive observations) and penalization $\lambda=h^{0.8}$ over the time interval $[0,72]$. The trajectory has been subsampled to $1$ minute between observation times. The initial position and velocity are set to $U_0=\begin{pmatrix} 15 & 20 & 0 & 0\end{pmatrix}^\top$.
The parameters of the SDE are $\tau=1$, $\nu=5$ and $\omega=0.1$. The weights in the potential are $\alpha_1=70$ and $\alpha_2=50$. The precision matrices $B_1$ and $B_2$ are chosen as 
\begin{equation}B_1=\begin{pmatrix} \frac{1}{9} & \frac{1}{40} \\  \frac{1}{40} & \frac{1}{4} \end{pmatrix} ; \quad B_2=\begin{pmatrix} \frac{1}{36} & -\frac{1}{100} \\  -\frac{1}{100} & \frac{1}{100} \end{pmatrix}. 
	\label{eq: precision matrices values}
\end{equation}

On average, the animal spends more time near the potential centers $x_1^*$ and $x_2^*$, which represent the preferred habitat. The movement is persistent in time due to the autocorrelation parameter $\tau>0$, and areas outside the domain $\mathcal{D}$ are very rarely visited as the penalization reorients the velocity inward each time the position falls outside the boundaries of $\mathcal{D}$. The model thus captures key features of wildlife movement: directional persistence, attraction toward biologically meaningful sites such as feeding or mating areas, and confinement to the appropriate habitat. This combination produces a realistic and interpretable movement model consistent with many ecological applications.

\section{Filtering algorithms}
\label{sec: filtering algorithms}
In this section, we present algorithms for estimating the latent states
$U_j = \begin{pmatrix} X_j^\top & V_j^\top \end{pmatrix}^\top$ of the
movement process from noisy observations $y_j$, $j \in \{1;\cdots,n\}$.
We first introduce the two main filtering methods (Kalman filter
and the particle filter) and then detail their implementation in our
specific setting.
The Extended Kalman Filter (Section~\ref{sec: ekf}) is used when the
measurement errors are Gaussian, exploiting the Gaussian
structure of the Lie-Trotter approximation.
For non-Gaussian measurement errors such as Student's-t or the
X-shaped Argos errors, we resort to particle filtering
(Sections~\ref{sec: pf student} and~\ref{sec: pf argos}), and develop
algorithms based on both the Lie-Trotter and Strang splitting schemes.
 
\subsection{Kalman filter and particle filter}
 
\paragraph{Kalman filter.}
The Kalman filter \citep{kalman_general_1960} is the optimal linear filter
for Gaussian state-space models: given a linear dynamic and Gaussian
noise, it recursively computes the exact filtering distribution
$p(U_j \mid Y_{1:j})$, which is Gaussian at every step.
When the dynamics are nonlinear, the Extended Kalman Filter (EKF)
linearizes the transition function around the current estimate using its Jacobian, yielding an approximate Gaussian filter that remains computationally efficient.
 
\paragraph{Particle filter.}
When the state-space model is nonlinear or the noise is non-Gaussian,
the filtering distribution $p(U_j \mid Y_{1:j})$ is no longer exact.
Particle filters, also called Sequential Monte Carlo (SMC) methods
\citep{doucet_tutorial_2011}, approximate this distribution by a
weighted set of $K$ particles $\{U_j^{(k)}, W_j^{(k)}\}_{k=1}^K$.
At each time step, particles are propagated through a proposal
distribution $q(U_j \mid U_{j-1}, y_j)$ and reweighted according to
how well they explain the new observation.
The choice of proposal distribution affects the efficiency
of the algorithm: the optimal proposal $p(U_j \mid U_{j-1}, y_j)$
minimises the variance of the weights \citep{doucet_tutorial_2011},
but is rarely available in closed form and must be approximated.
The generic SMC algorithm is given in Algorithm~\ref{algo: particle filter}.
 
\begin{algorithm}[H]
	\caption{SMC algorithm}\label{algo: particle filter}
	\begin{algorithmic}
		\renewcommand{\algorithmicrequire}{\textbf{Input:}}
		\renewcommand{\algorithmicensure}{\textbf{Output:}}
		\Require Data $y_{1:n}$, number of particles $K$, proposal $q$,
		         initial distribution $\pi_0$
		\State Sample $U_0^{(k)}  \sim \pi_0(\cdot)$ for
		       $k \in \{1,\ldots,K\}$.
		\State Compute and normalize the weights:
		\[\omega_0^{(k)}=p(y_0,U_0^{(k)}); \quad
		  W_0^{(k)}=\frac{\omega_0^{(k)}}{\sum_{k=1}^K \omega_0^{(k)}}.\]
		\For{$j=1,\ldots,n$}
		\For{$k=1,\ldots,K$}
		\State Resample $U^{\prime(k)}_{0:j-1}$ from
		       $\Psi^{K}_{j-1}=\sum_{k=1}^K W_{j-1}^{(k)}
		       \delta_{U^{\prime(k)}_{0:j-1}}$.
		\State Propagate: sample
		       $U_j^{(k)}\sim q(\cdot \mid y_j,U^{\prime(k)}_{j-1})$.
		\State Compute and normalize the importance weights:
		\[w_j^{(k)}=\frac{p(y_j \mid U_j^{(k)})
		  p(U_j^{(k)} \mid U^{\prime(k)}_{j-1})}
		  {q(U_j^{(k)}\mid y_j,U^{\prime(k)}_{j-1})}; \quad
		  W_j^{(k)}=\frac{w_j^{(k)}}{\sum_{k=1}^K w_j^{(k)}}.\]
		\EndFor
		\EndFor
		\Ensure Weighted particles $\{U_j^{(k)}, W_j^{(k)}\}_{k=1}^K$
		        approximating $p(U_j \mid Y_{1:j})$ for each $j$.
	\end{algorithmic}
\end{algorithm}
 
After running the algorithm, the conditional mean
$\mathbb{E}(X_j \mid Y_{1:j})$ is approximated by
\[
\hat{\mathbb{E}}(X_j \mid Y_{1:j})
= \sum_{k=1}^K W_{j}^{(k)} X_j^{(k)}.
\]
In the following sections, we specify the proposal distribution $q$
for each combination of measurement error model and splitting scheme.

\subsection{Extended Kalman filter for Gaussian measurement errors}
\label{sec: ekf}
We assume that the true positions $X_j=X(jh)$, $j \in \{1,\ldots,n\}$ are observed with Gaussian measurement noise. We use the Lie-Trotter approximation of the state process to obtain a Gaussian state-space model for subsequent Kalman filtering. The non-linear Gaussian state-space is then
\begin{align}
	\begin{cases}
		U^{[\mathrm{LT}]}_{j+1}=e^{\tilde{A}h}U^{[\mathrm{LT}]}_j-e^{\tilde{A}h}u_l^*-he^{\tilde{A}h}g_{\lambda}(U^{[\mathrm{LT}]}_j))+u_l^*+\eta_j \qquad \eta_j \sim \mathcal{N}(0_{4,1},\tilde{Q})  \\
		y_j=LU^{[\mathrm{LT}]}_j+\varepsilon_j \qquad \varepsilon_j \sim \mathcal{N}(0_{2,1},\sigma_{obs}^2I_2)
	\end{cases}
\label{eq: LT state space}
\end{align}
where $L=\begin{pmatrix} I_2 & 0_2\end{pmatrix}$ and the other matrices are as before. The subscript $h$ is suppressed for convenience. The non-linearity comes from the potential $H$ as well as the penalization $\beta_\lambda$ encoded in the function $g_{\lambda}$.
We now detail the Extended Kalman Filter algorithm that produces state estimates $\hat{U}_j^{[\mathrm{LT}]}$, $j \in\{1,\ldots,n\}$ approximating both the position and the velocity given the observations.

First, initialize the vector of positions and velocities $\hat{U}_0$ and their covariance $R_0$. Then iterate as follows:
\begin{align}
	&\hat{U}_{j^-}^{[\mathrm{LT}]}=e^{\tilde{A}h}\hat{U}_{j-1}^{[\mathrm{LT}]}-e^{\tilde{A}h}u_l^*-he^{\tilde{A}h}g_{\lambda}(\hat{U}_{j-1}^{[\mathrm{LT}]})+u_l^* ,\\
	&R_{j^-}^{[\mathrm{LT}]}=e^{\tilde{A}h}R_{j-1}^{[\mathrm{LT}]}e^{\tilde{A}^\top h}+h^2e^{\tilde{A}h}G_{j-1} R_{j-1}^{[\mathrm{LT}]} G_{j-1}^\top e^{\tilde{A}^\top h }+\tilde{Q},
\end{align}
where $G_{j-1}=Dg_{\lambda}(\hat{U}_{j-1}^{[\mathrm{LT}]})=\left( \frac{\partial g_{\lambda}^{(i)}}{\partial u_k}(\hat{U}_{j-1}^{[\mathrm{LT}]})\right)_{1\leq i,k\leq 4}$.

Let $g_{v,\lambda}$ denote the velocity components for the function $g_{v,\lambda}$, that is 
\[g_{v,\lambda}(x,v)=2\alpha_l(e_l(x)-1)B_l(x-x^*_l)+\nabla H_{-l}(x)+\beta_\lambda(x). \]  \[G_{j-1}=\begin{pmatrix} 0_{2} & 0_2 \\ D_xg_{v,\lambda}(\hat{U}_{j-1}^{[\mathrm{LT}]}) & 0_2\end{pmatrix}\]
where $D_x$ is the Jacobian matrix with respect to the $x$ components,
$$D_xg_{v,\lambda}(x)=\frac{I_2-D\pi(\hat{X}_{j-1}^{[\mathrm{LT}]})}{\lambda}+D^2 H_{-l}(x)+2\alpha_l\left((e_l(x)-1)B_l-2e_l(x)B_l(x-x^*_l)(x-x^*_l)^\top B_l\right),$$
where $D\pi$ is the Jacobian matrix of the projection operator and the Hessian of the potential surface is given by \eqref{eq: potential hessian}.
Since the projection $\pi$ is piecewise linear in $x$, the Jacobian is  
\[D\pi(x)=\frac{(v_{k+1}-v_k)(v_{k+1}-v_k)^\top}{l^2}.\]
Thus, the correction step of the extended Kalman filter is 
\begin{align}
	&K_j^{[\mathrm{LT}]}=R_{j^-}^{[\mathrm{LT}]}L^\top(LR_{j^-}^{[\mathrm{LT}]}L^\top+\sigma_{obs}^2I_2)^{-1},\\
	&\hat{U}_j^{[\mathrm{LT}]}=\hat{U}_{j^-}^{[\mathrm{LT}]}+K_j^{[\mathrm{LT}]}(y_j-L\hat{U}_{j^-}^{[\mathrm{LT}]}),\\
	&R_j^{[\mathrm{LT}]}=(I_4-K_j^{[\mathrm{LT}]}L)R_{j^-}^{[\mathrm{LT}]}.
\end{align}

\subsection{Particle filter for Student's t-distribution measurement errors}
\label{sec: pf student}
Observed animal positions often suffer from non-Gaussian measurement errors, for example for Fastloc-GPS observations \citep{wensveen_path_2015}, which are better described by a Student's t-distribution. Then Kalman filtering techniques may be unreliable, especially if the error distribution differs a lot from a Gaussian distribution. To address this, we resort to particle filtering techniques. We develop two particle filtering algorithms depending on the splitting scheme (Lie-Trotter or Strang), which is used to approximate the solution to the SDE. We compare their performances in Section~\ref{sec: simulation study}.

\subsubsection{Particle filter based on Lie-Trotter scheme}

The state-space updates derived from the Lie-Trotter approximation of the SDE are:
\begin{align*}
\begin{cases}
	U^{[\mathrm{LT}]}_{j+1}=e^{\tilde{A}h}U^{[\mathrm{LT}]}_j-e^{\tilde{A}h}u_l^*-he^{\tilde{A}h}g_{\lambda}(U^{[\mathrm{LT}]}_j)+u_l^*+\eta_j; \quad \eta_j \sim \mathcal{N}(0_{4,1},\tilde{Q}),  \\
	y_j=LU^{[\mathrm{LT}]}_j+\sigma_{obs}\varepsilon_j; \quad \varepsilon_j \sim t_d.
\end{cases}
\end{align*}
The Student's t-distribution can be approximated by a Gaussian distribution with standard deviation $\sigma_{obs}'=\frac{d-2}{d}\sigma_{obs}$ to transform the equations into a Gaussian state-space model and then resort to Kalman filtering. 

However, without this simplification, the filtering distributions $p(U_j \vert Y_{1:j})$ are not explicit and a particle filter is needed to approximate it. We propose to approximate the optimal proposal as follows.  Bayes's formula gives 
\[p(U_j \vert U_{j-1},y_j) \propto p(U_j\vert U_{j-1})p(y_j \vert U_j).\]
We replace $p(U_j\vert U_{j-1})$ in this equation by an approximation  $p^{[\mathrm{LT}]}(U_j\vert U_{j-1})$ obtained from the Lie-Trotter splitting scheme. We approximate $p(y_j \vert U_j)$ in the proposal by a Gaussian density $\tilde{p}(y_j \vert U_j)$ with mean $LU_j$ and covariance matrix $\sigma_{obs}^2\frac{d}{d-2} I_2$. Hence, the new proposal $q(U_j \vert U_{j-1},y_j) \propto p^{[\mathrm{LT}]}(U_j\vert U_{j-1})\tilde{p}(y_j \vert U_j)$ is a product of Gaussian densities. It is therefore a Gaussian proposal with covariance matrix and mean given by
\[\Gamma^{[\mathrm{LT}]}=\left(\tilde{Q}^{-1}+\frac{1}{\sigma_{obs}^2} \frac{d-2}{d} L^\top L\right)^{-1},\]
\[m_j^{[\mathrm{LT}]}=\Gamma^{[\mathrm{LT}]} \left( \tilde{Q}^{-1}(e^{\tilde{A}h}(U_{j-1}-u^*_l-hg_{\lambda}(U_{j-1}))+u^*_l)+\frac{1}{\sigma_{obs}^2} \frac{d-2}{d}L^\top y_j\right).\]
It is then easy to propagate the particles using the Gaussian proposal. The weights are also easily computed as follows:
\[w_j^{(k)}=\frac{p(y_j \vert U_j^{(k)})p^{[\mathrm{LT}]}(U_j^{(k)} \vert U^{'(k)}_{j-1})}{\mathcal{N}(U_j^{(k)};m_j^{[\mathrm{LT}]},\Gamma^{[\mathrm{LT}]})}\]
where $\mathcal{N}(U_j^{(k)};m_j^{[\mathrm{LT}]},\Gamma^{[\mathrm{LT}]})$ is the density of a Gaussian variable with mean $m_j^{[\mathrm{LT}]}$ and covariance $\Gamma^{[\mathrm{LT}]}$ evaluated at $U_j^{(k)}$.

In this algorithm, the penalization plays a role in the propagation of the particles as well as in the weighting of each particle through the Gaussian proposal.


\subsubsection{Particle filter based on the Strang approximation}
We now introduce a particle filter based on the Strang splitting scheme.
The optimal proposal in the particle filter can be rewritten as 
\begin{equation*}p(U_j\vert U_{j-1},y_j)\propto p(X_j\vert U_{j-1})p(V_j \vert X_j,U_{j-1}) p(y_j \vert U_j).
\end{equation*}
We define the proposal distribution as
\begin{equation}q(U_j\vert U_{j-1},y_j)\propto p^{[\mathrm{S}]}(V_j \vert X_j,U_{j-1}) p^{[\mathrm{S}]}(X_j\vert U_{j-1}) \tilde{p}(y_j \vert U_j)
\label{eq : Strang proposal up to multiplicative constant}\end{equation}
where $\tilde{p}(y_j \vert U_j)$ is a Gaussian density with mean  $LU_j$ and covariance matrix $\sigma_{obs}^2\frac{d}{d-2} I_2$.
Hence, the product $p^{[\mathrm{S}]}(X_j\vert U_{j-1}) \tilde{p}(y_j \vert U_j)$ is Gaussian with covariance matrix and mean
\[\Gamma^{[\mathrm{S}]}=(\tilde{Q}_{xx}^{-1}+\frac{1}{\sigma_{obs}^2}\frac{d-2}{d}I_2)^{-1};\quad \hat{x}^{[\mathrm{S}]}_j=\Gamma^{[\mathrm{S}]}(\tilde{Q}_{xx}^{-1}\tilde{m}_x+\frac{1}{\sigma_{obs}^2}\frac{d-2}{d}y_j).
\\\]
Note that $\tilde{p}(y_j \vert U_j)=\tilde{p}(y_j \vert X_j)$ so that the normalisation constant in \eqref{eq : Strang proposal up to multiplicative constant} is given by $\int p(X_j \vert U_{j-1}) \tilde{p}(y_j \vert X_j) dX_j$ and we obtain the proposal
\[q(U_j\vert U_{j-1},y_j) = p^{[\mathrm{S}]}(V_j \vert X_j,U_{j-1}) \mathcal{N}(X_j ; \hat{x}^{[\mathrm{S}]}_j,\Gamma^{[\mathrm{S}]}).\]
The propagation of the particles is thus done in two steps. First, we propagate the position based on the current position and velocity with a Gaussian kernel. Then we propagate the velocity given the new position. As mentioned in section~\ref{sec: splitting schemes}, $p^{[\mathrm{S}]}(V_j \vert X_j,U_{j-1})$ is explicit and Gaussian.
This produces a simple formula for the updates of weights 
\[w_j^{(k)}=\frac{p(y_j\vert U_j^{(k)}) p^{[\mathrm{S}]}(X_j^{(k)}\vert U^{'(k)}_{j-1})}{\mathcal{N}(X_j^{(k)};\hat{x}^{[\mathrm{S}]}_j,\Gamma^{[\mathrm{S}]})}.\]

\subsection{Particle filter for X-shaped measurement errors}
\label{sec: pf argos}
For Argos telemetry data, we here present a particle filter algorithm, based on either the Lie-Trotter or the Strang approximation.

\subsubsection{Particle filter based on the Lie-Trotter approximation}

For the proposal in the SMC algorithm, we approximate the Bayes formula defining the posterior distribution.  We consider a mixture $\tilde{p}(y_j \vert U_j)$ of Gaussian densities with covariance matrices $\Sigma$ and $\tilde{\Sigma}$ to approximate the term $p(y_j \vert U_j)$.
For Lie-Trotter, $p^{[\mathrm{LT}]}(U_j\vert U_{j-1})\tilde{p}(y_j \vert U_j)$ is a product of Gaussian densities. Hence, the proposal $q$ is a mixture of two multivariate Gaussians, which is easy to simulate. The covariance matrices are
\[\Gamma^{[\mathrm{LT}]}=(\tilde{Q}^{-1}+\frac{d-2}{d}L^\top\Sigma L)^{-1}; \quad \tilde{\Gamma}^{[\mathrm{LT}]}=(\tilde{Q}^{-1}+\frac{d-2}{d}L^\top\tilde{\Sigma} L)^{-1} \]
and the means are
\begin{align}&m_j^{[\mathrm{LT}]}=\Gamma^{[\mathrm{LT}]}(\tilde{Q}^{-1}(e^{\tilde{A}h}(U_{j-1}-u^*_l-hg_{\lambda}(U_{j-1}))+u^*_l)+\frac{d-2}{d}L^\top \Sigma^{-1} y)), \\ &\tilde{m}_j^{[\mathrm{LT}]}=\tilde{\Gamma}^{[\mathrm{LT}]}(\tilde{Q}^{-1}(e^{\tilde{A}h}(U_{j-1}-u^*_l-hg_{\lambda}(U_{j-1}))+u^*_l)+\frac{d-2}{d}L^\top \tilde{\Sigma}^{-1} y)).
\end{align}
The update of the weights is then:
\[w_j^{(k)}=\frac{p(y_j\vert U_j^{(k)}) p^{[\mathrm{LT}]}(U_j^{(k)}\vert U_{j-1}^{'(k)})}{p\mathcal{N}(U_j^{(k)};m_j^{[\mathrm{LT}]},\Gamma^{[\mathrm{LT}]})+(1-p)\mathcal{N}(U_j^{(k)};\tilde{m}^{[\mathrm{LT}]}_j,\tilde{\Gamma}^{[\mathrm{LT}]})}.\]

\subsubsection{Particle filter based on the Strang approximation}
Similarly, we approximate $p(y_j\vert U_j)$ in the proposal by a mixture of Gaussian densities with covariance matrices $\Sigma$, $\tilde{\Sigma}$.
The product $p^{[\mathrm{S}]}(X_j \vert U_{j-1})\tilde{p}(y_j\vert U_j)$ is thus a mixture of Gaussian densities with covariance matrices 
\[\Gamma^{[\mathrm{S}]}=(Q_{xx}^{-1}+\frac{d-2}{d}\Sigma^{-1})^{-1}; \quad \tilde{\Gamma}^S=(Q_{xx}^{-1}+ \frac{d-2}{d}\tilde{\Sigma}^{-1})^{-1}\]
and means 
\[\hat{x}_j^{[\mathrm{S}]}= \Gamma^{[\mathrm{S}]}(\tilde{Q}_{xx}^{-1}\tilde{m}_x+\frac{d-2}{d}\Sigma^{-1} y_j) \quad \tilde{\hat{x}}_j^{[\mathrm{S}]}=\tilde{\Gamma}^{[\mathrm{S}]}(\tilde{Q}_{xx}^{-1}\tilde{m}_x+\frac{d-2}{d}\tilde{\Sigma}^{-1}y_j).\]
The update of the weights is 
\[w_j^{(k)}=\frac{p(y_j\vert U_j^{(k)}) p^{[\mathrm{S}]}(X_j^{(k)}\vert U^{'(k)}_{j-1})}{p\mathcal{N}(X_j^{(k)};\hat{x}^{[\mathrm{S}]}_j,\Gamma^{[\mathrm{S}]})+(1-p)\mathcal{N}(X_j^{(k)};\tilde{\hat{x}}^{[\mathrm{S}]}_j,\tilde{\Gamma}^{[\mathrm{S}]})}.\]

\section{Simulation study}
\label{sec: simulation study}

Kalman and Extended Kalman filters are implemented in \texttt{R}. For efficiency, the particle filter is implemented in \texttt{C++} and called from \texttt{R} via the \texttt{Rcpp} package \citep{eddelbuettel_rcpp_2011}.
We start by evaluating the filtering methods in a standard scenario where measurement errors are Gaussian. This setting allows to isolate the effects of domain penalization and sampling frequency on the filtering accuracy. We then assess the performance of particle filters for the non-Gaussian, heavy-tailed error models introduced in Section \ref{sec: measurement error model}, and compare their accuracy with that of the Extended Kalman filter, in which the non-Gaussian errors are approximated by Gaussian errors with matching variances. The domain $\mathcal{D}$ is a polygon with $19$ vertices (see Figure~\ref{fig:trajectory mix gaussian potential}).
The values of the movement parameters and the potential parameters are given at the end of Subsection \ref{subsec: approximate path sim}.

\subsection{Gaussian errors}

We simulate $50$ trajectories over a period $[0,T]$ with $T=72$ hours and constant time step $h=1$ second based on the Lie-Trotter approximation. The penalization is fixed to $\lambda=h^{0.8}$. We add Gaussian measurement errors with standard deviation $\sigma_{obs}=0.2$ to create noisy observations of the trajectories.

The data is then subsampled to time steps $1$ min, $3$ min, $5$ min and $20$ min.  We test the filtering algorithms for the four sampling intervals.
 
We compare the results of four methods:
 1) Kalman filtering without domain penalization ($\lambda=+\infty$) where the non-linear part in the state space is treated as a constant term (KF); 2)  Extended Kalman filtering without domain penalization ($\lambda=+\infty$) (EKF); 3) Kalman filtering with domain penalization (Penalized KF) and 4) Extended Kalman filtering with domain penalization (Penalized EKF).

\begin{figure}[ht!]
	\centering
	\begin{subfigure}{0.8\linewidth}
		\centering
		\includegraphics[width=0.8\linewidth]{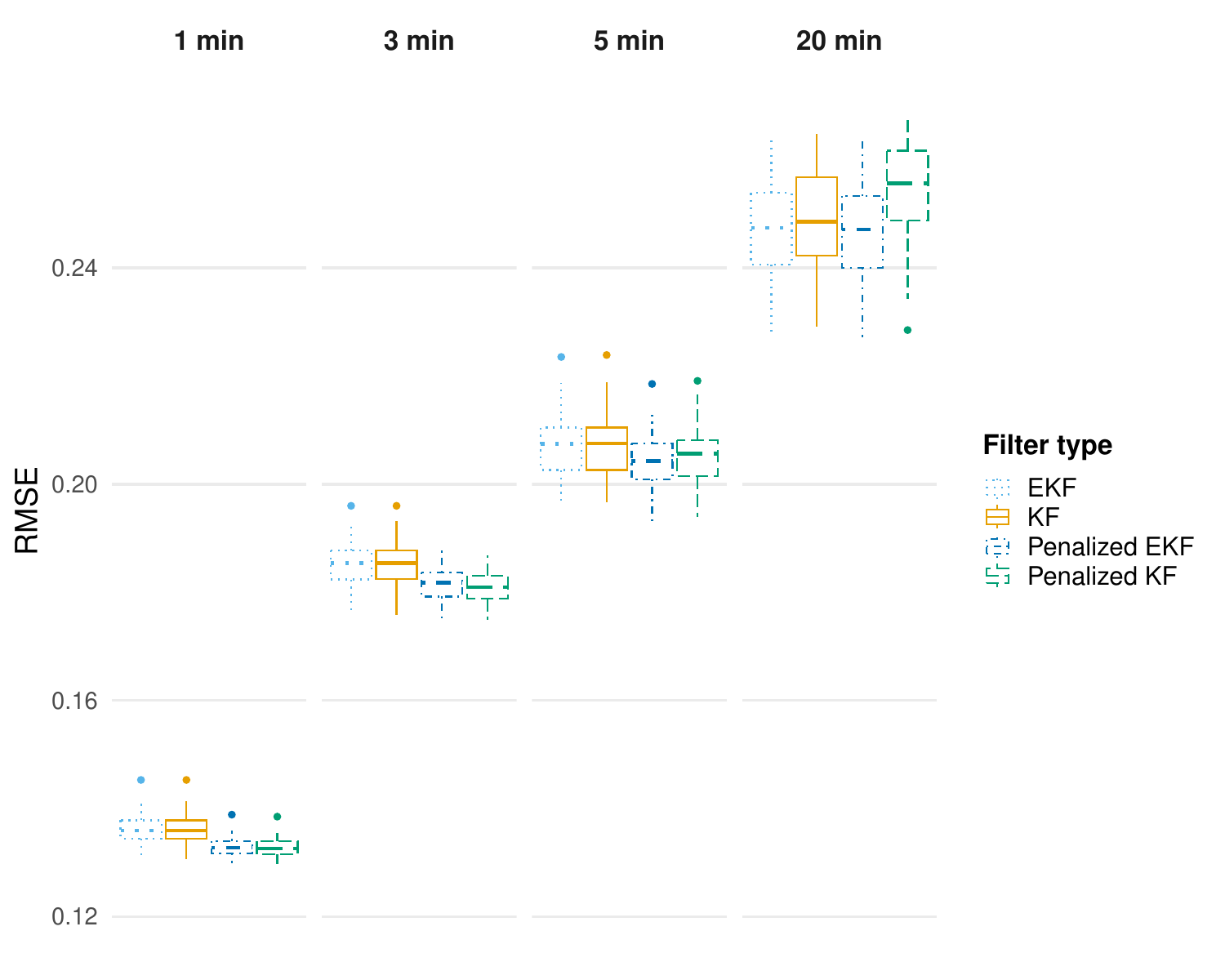}
		\caption{}
		\label{fig:ekfkrmse}
	\end{subfigure}
	
	\begin{subfigure}{0.8\linewidth}
		\centering
		\includegraphics[width=0.8\linewidth]{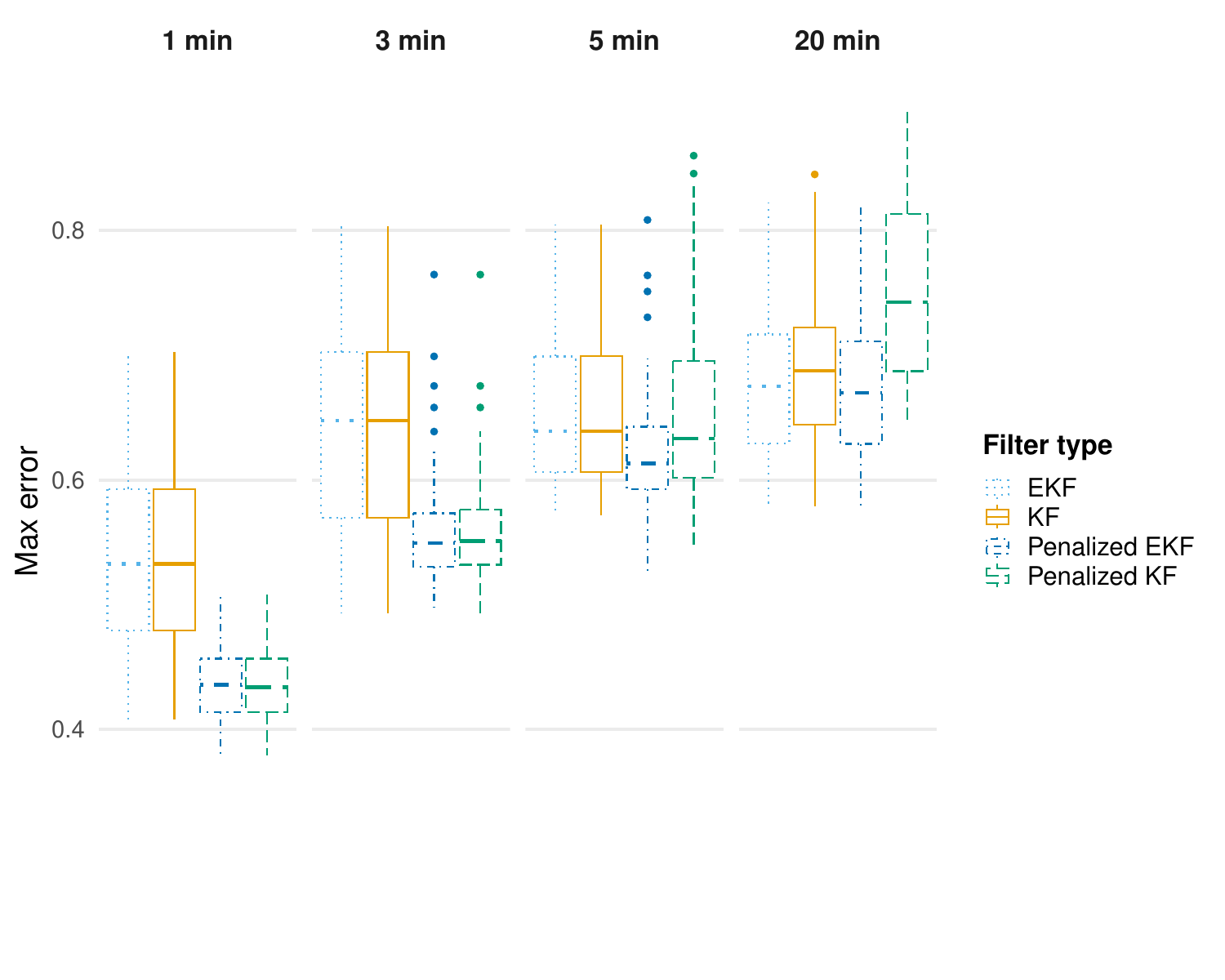}
		\caption{}
		\label{fig:ekfkfmax}
	\end{subfigure}
  
	\caption{Results of filtering for $50$ simulated trajectories with additive Gaussian noise. (a) Boxplot of RMSE for KF, EKF, Penalized KF and Penalized EKF. (b) Boxplot of max errors. }
	\label{fig:ekfkferrors}
\end{figure}

The results are shown in Figure~\ref{fig:ekfkferrors}. Figure~\ref{fig:ekfkrmse} shows boxplots of the root mean squared error (RMSE) for each value of the time step $h$ using KF, EKF, Penalized KF and Penalized EKF. The theoretical RMSE before filtering is equal to $\sqrt{\frac{\pi}{2}} \sigma_{obs}\simeq 0.25$. The incorporation of spatial constraints through the domain penalization reduces the RMSE for small time steps. For a time step of $h=20$ minutes, the Kalman filters with or without domain penalization have slightly worse performance than the extended Kalman filters. The reason is that the approximation of constant potential over each time step that is used in the naive Kalman filter is very poor for large time steps. The extended Kalman filter then provides a slightly better approximation of the potential surface and hence slightly better accuracy.

Figure~\ref{fig:ekfkfmax} shows boxplots of the maximum error between the true positions and the filtered positions on the $50$ trajectories for each value of the time step $h$ with the four filters. The incorporation of domain knowledge improves the performance of the filter for high frequency observations ($h=1, 3$ or $5$ min). Again, the Kalman filter has slightly worse performance than the Extended Kalman filter for large time steps ($h=20$ min).
Mean values of the maximum error and RMSEs over the $50$ trajectories are shown in Table~\ref{tab:ekfkferrors}. The simulation study was performed on a personal laptop and took $30$ minutes to run.

 Overall, the penalized filters avoid large errors in the reconstructed trajectories and ensure that they remain within or close to the spatial domain of interest. The benefits of domain penalization are most pronounced for small time steps (high-frequency observations), where the filter has enough temporal resolution to adjust to the local shape of the domain. 

\begin{table}[ht!]
	\centering
	\begin{tabular}{|cccccc|}
		\hline
		$h$ & KF & EKF &  Penalized KF & Penalized EKF & Before filter\\
		\hline
		& \multicolumn{5}{c|}{Average max error (km)} \\
		\hline
		$1$ min  & 0.540 & 0.540  & {\bf 0.437} & {\bf 0.437} & 0.851\\
		$3$ min & 0.643 & 0.643 & {\bf 0.561} & 0.562 & 0.792\\
		$5$ min & 0.656 & 0.656 & 0.657 & {\bf 0.626} & 0.761\\
		$20$ min & 0.694 & 0.682 &  1.087 & {\bf 0.680} & 0.686\\
		\hline
		& \multicolumn{5}{c|}{Average RMSE (km)} \\
		\hline
		$1$ min  & 0.136  & 0.136  & {\bf 0.133} & {\bf 0.133} & 0.251\\
		$3$ min & 0.185 & 0.185 & 0.182 & {\bf 0.181} & 0.250\\
		$5$ min & 0.207& 0.207& 0.205 & {\bf 0.204} & 0.251\\
		$20$ min & 0.249 & {\bf 0.247} &  0.256 & {\bf 0.247} & 0.250\\
		\hline
	\end{tabular}
	\caption{Max error and root mean squared error (RMSE) averaged over the $50$ trajectories with Gaussian measurement error for different values of the time step $h$. The smallest value in each row is marked in bold.}
	\label{tab:ekfkferrors}
\end{table}

\subsection{Student's t-distributed errors}

We simulate $20$ trajectories over a period $[0,T]$ with $T=48$ based on the Strang approximation. The penalization parameter is again set to $\lambda=h^{0.8}$, where $h=\frac{1}{3600}$ denotes the simulation time step. The data is subsampled as previously. We add $t_d$ measurement errors with scale parameter $\sigma_{obs}=0.2$  and degrees of freedom $d=3$ to create noisy observations of the trajectories.
We run an EKF and a Penalized EKF where the $t_d$ is approximated by a Gaussian distribution with variance $\frac{d}{d-2}\sigma_{obs}^2$. Then, we run a particle filter without domain penalization based on the Lie-Trotter  scheme (Lie-Trotter PF), a particle filter with domain penalization based on the Lie-Trotter scheme  (Penalized Lie-Trotter PF), a particle filter without domain penalization based on the Strang scheme (Strang PF) and a particle filter with domain penalization based on the Strang scheme (Penalized Strang PF). We choose $K=1000$ particles in all cases. The calculations are performed in parallel on a remote cluster over a single node of $48$ cores. The particle filters are longer to run than the Kalman filters, even with the \texttt{C++} implementation. The whole simulation took about $1$ hour.

\begin{figure}[ht!]
	\centering
	\begin{subfigure}{0.9\linewidth}
		\centering
		\includegraphics[width=0.8\linewidth]{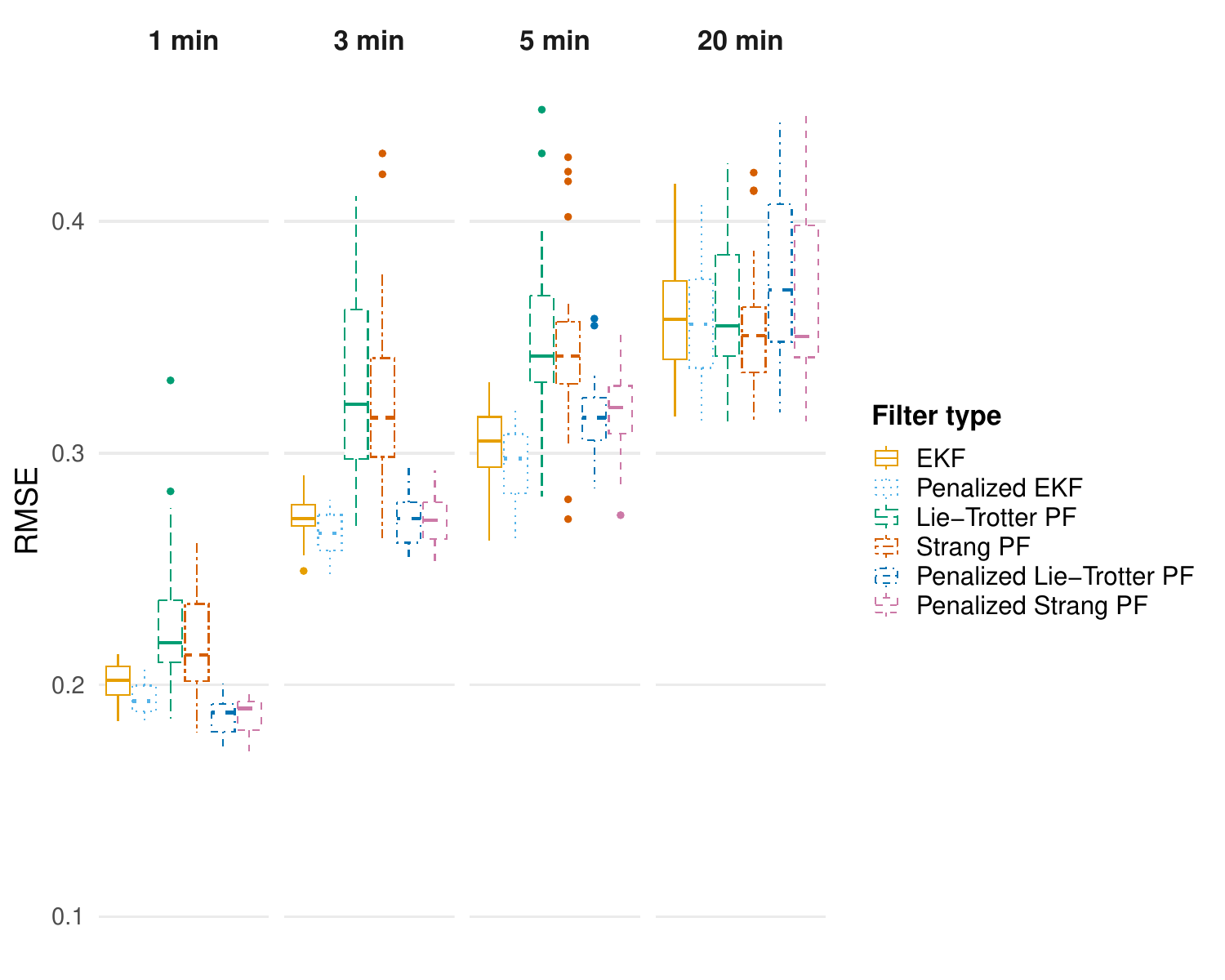}
		\caption{}
		\label{fig:studentrmse}
	\end{subfigure}
	
	\begin{subfigure}{0.9\linewidth}
		\centering
		\includegraphics[width=0.8\linewidth]{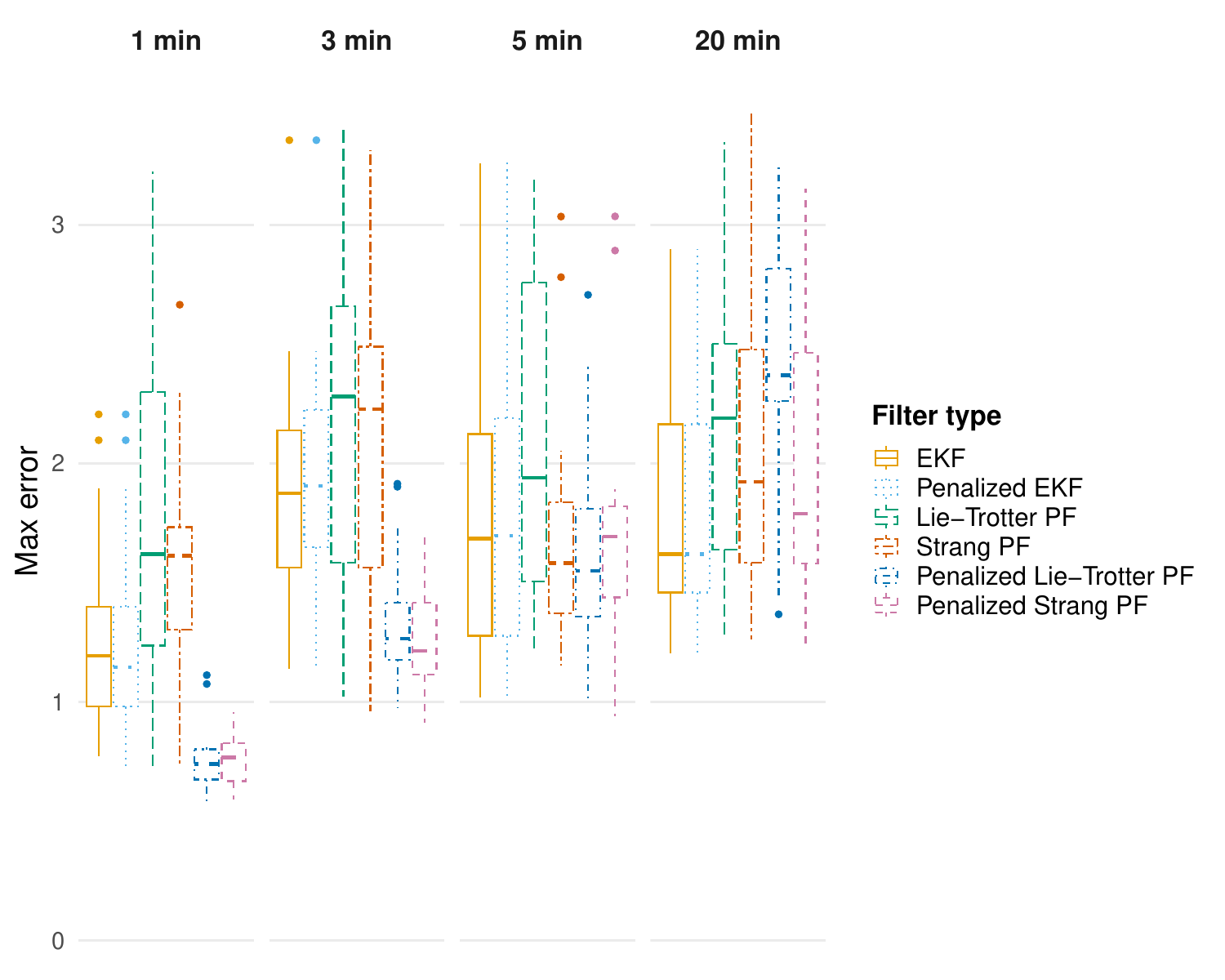}
		\caption{}
		\label{fig:studentmax}
	\end{subfigure}
    
    \caption{Results of filtering for $20$ simulated trajectories with additive Student noise.
    (a) Boxplots of root mean squared error (RMSE) for EKF, Penalized EKF, Strang PF, Penalized Strang PF , Lie-Trotter PF, Penalized Lie-Trotter PF. (b) Boxplots of max error. }
\end{figure}

Figure~\ref{fig:studentrmse} shows boxplots of the RMSE for each value of the time step $h$. The incorporation of domain knowledge associated with a particle filter to handle non-Gaussian error improves the performance of the filter for time steps $h=1$, $3$ and $5$ min. Notably, the RMSE for the particle filters without domain penalization is higher than the RMSE obtained with EKF. The misspecification of the underlying dynamics in the particle filter likely produces trajectories that progressively drift away from the true track each time we linger outside the boundary.
Figure~\ref{fig:studentmax} shows the same boxplots for the maximum error. The incorporation of domain knowledge also increases the performance of the filter for time steps $h=1$, $3$ and $5$ min. For $h=1$ min, the maximum error is reduced by about $40\%$ for the penalized particle filters compared to the EKF and penalized EKF.

From a practical perspective this demonstrates that including even simple geometric information about the movement domain can improve the reliability of track reconstruction from noisy telemetry data, particularly when measurement errors are heavy-tailed.

Moreover, the results seem to suggest that for the specific potential function that we chose here, Strang and Lie-Trotter approximation give sensibly similar results. We would have expected Strang to perform better than Lie-Trotter, and to handle better coarse observations since it is known to have better statistical properties \citep{pilipovic_parameter_2024}, but the simulation study did not show such clear pattern.  

Mean values of the maximum error and RMSE over the $20$ trajectories are shown in Table~\ref{tab:studenterrors}.

\begin{table}[ht!]
	\centering
	\begin{tabular}{|ccp{4.5em}p{4.5em}p{4.5em}p{4.5em}p{4.5em}p{4em}|}
		\hline
		$h$ & EKF & Penalized EKF &  Lie-Trotter PF & Penalized Lie-Trotter PF & Strang PF & Penalized Strang PF & Before filter\\
		\hline
		& \multicolumn{7}{c|}{Average max error (km)} \\
		\hline
		$1$ min  & 1.287  & 1.278 & 2.304 & {\bf 0.754} & 1.793 & 0.757 & 5.657\\
		$3$ min & 2.010 & 2.076 & 2.708 & 1.334 & 2.592 & {\bf 1.245} & 4.694\\
		$5$ min & 1.980 & 2.037 & 2.335 & 1.879 & 2.334 & {\bf 1.711} & 3.587\\
		$20$ min & 2.193 & {\bf 2.188} &  2.387 & 2.767 & 2.303 & 2.522 & 2.582 \\
		\hline
		& \multicolumn{7}{c|}{Average RMSE (km)} \\
		\hline
		$1$ min  & 0.202 & 0.194  & 0.229 & {\bf 0.187} & 0.217 & {\bf 0.187} & 0.361 \\
		$3$ min & 0.272 & {\bf 0.264} & 0.328 & 0.272 & 0.323 & 0.271 & 0.362\\
		$5$ min & 0.305 & {\bf 0.297}  & 0.352 & 0.323 & 0.344 & 0.318 & 0.357\\
		$20$ min & 0.365 & 0.365 &  0.363 & 0.386 & {\bf 0.356} & 0.380 & 0.356\\
		\hline
	\end{tabular}
	\caption{Max error and root mean squared error (RMSE) averaged over the $20$ trajectories with Student's t-distributed measurement errors for different values of the time step $h$. The smallest value in each row is marked in bold. The last column shows the average error in the observations (before filtering) compared to the true positions.}
	\label{tab:studenterrors}
\end{table}

\subsection{Argos errors}

\begin{figure}[ht!]
	\centering
	\begin{subfigure}{0.9\linewidth}
		\centering
		\includegraphics[width=0.8\linewidth]{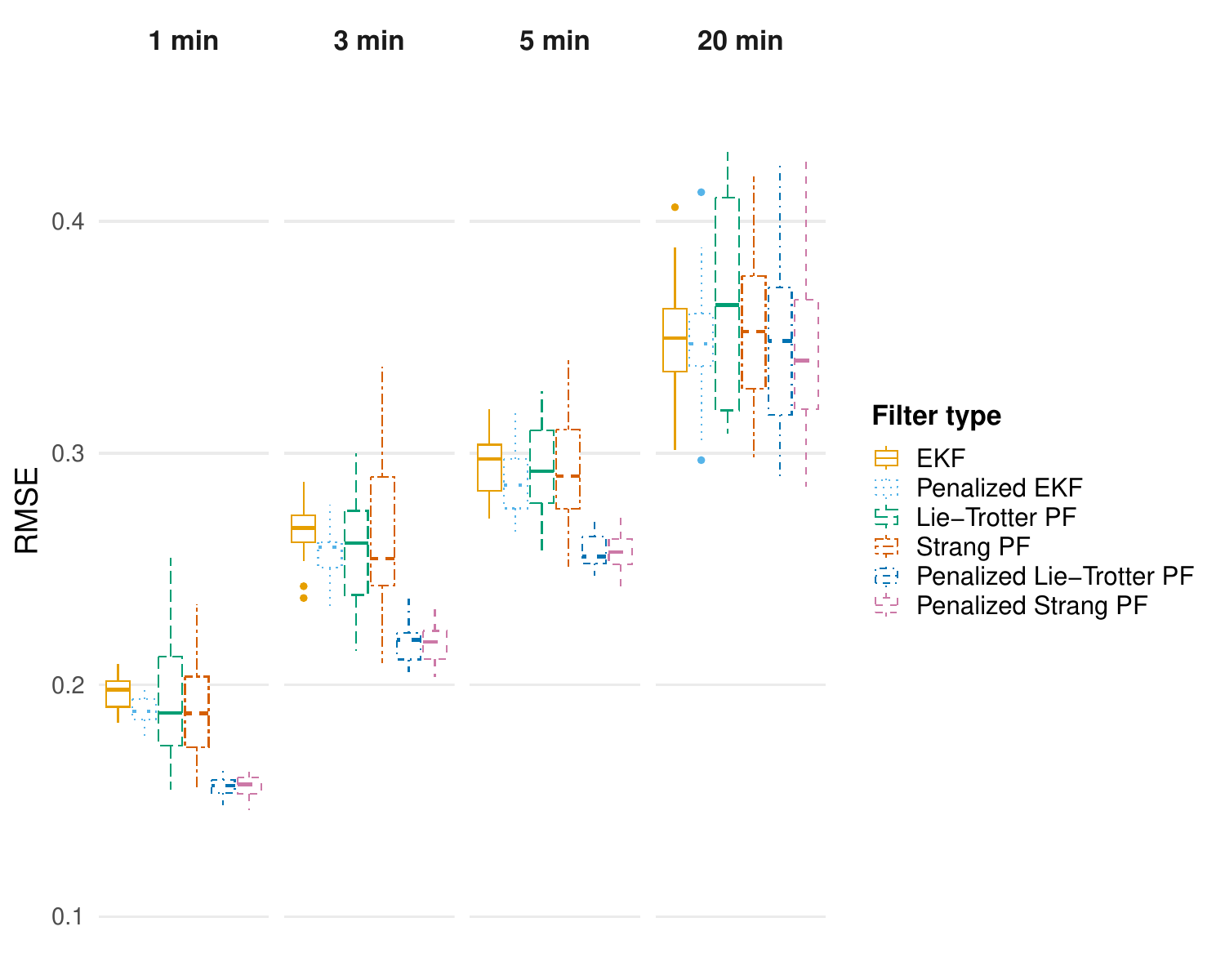}
		\caption{}
		\label{fig:argosrmse}
	\end{subfigure}
	
	\begin{subfigure}{0.9\linewidth}
		\centering
		\includegraphics[width=0.8\linewidth]{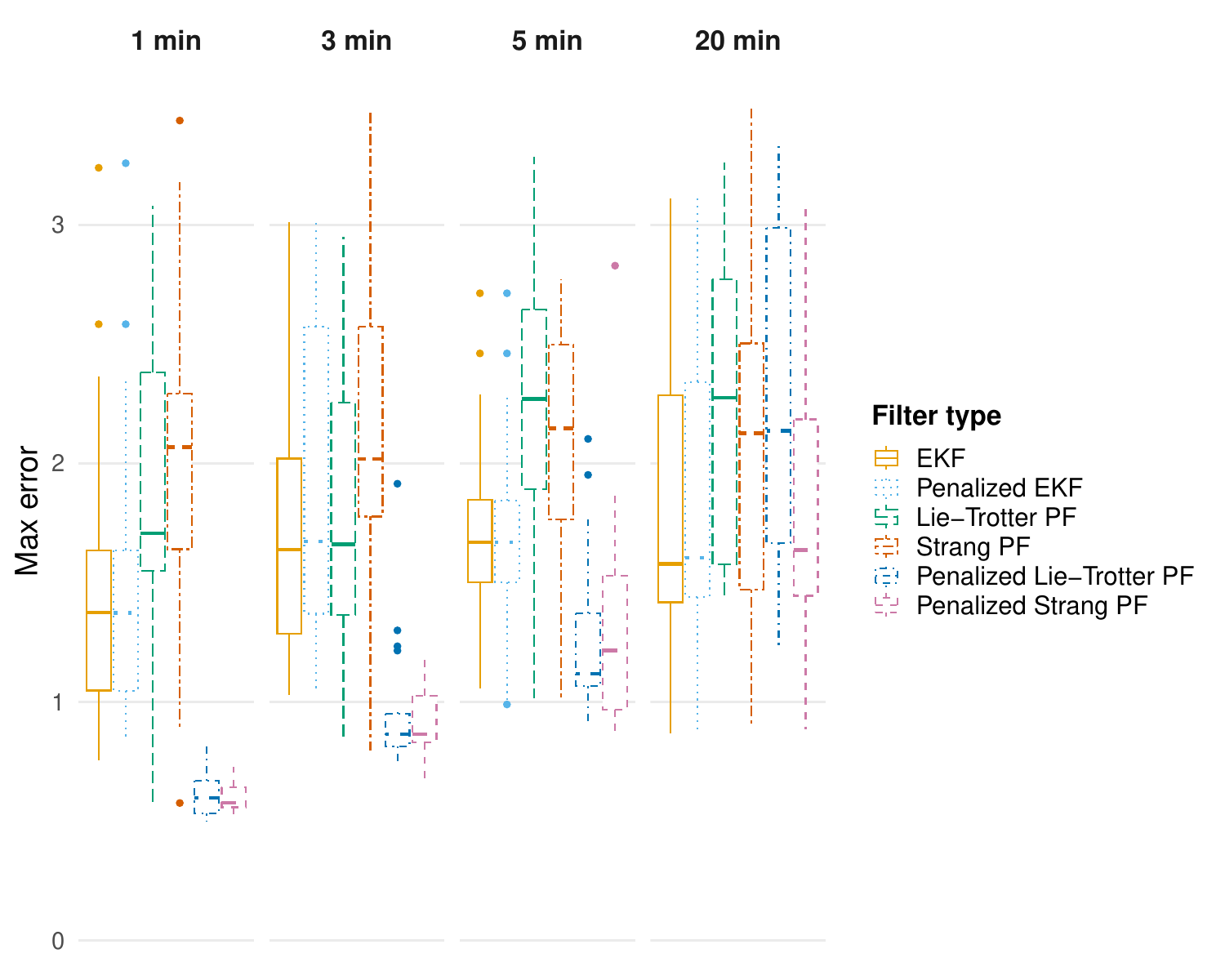}
		\caption{}
		\label{fig:argosmax}
	\end{subfigure}
    \caption{Results of filtering for $20$ simulated trajectories with additive Argos X-shaped noise. (a) Boxplots of RMSE (b) Boxplots of max error. } 
\end{figure}

We consider the same setup. We add X-shaped Argos measurement errors with scale parameters $\sigma_{obs}=0.2$, $\rho=0.7$, $a=0.4$, $p=0.5$, and degrees of freedom $d=3$ to create noisy observations of the trajectories. We run an EKF, a Penalized EKF for which the error distribution is approximated by an isotropic Gaussian distribution with variance $\frac{d}{d-2}\sigma_{obs}^2$. Then, we run a Lie-Trotter PF, a Penalized Lie-Trotter PF, a Strang PF and a Penalized Strang PF with $K=1000$ particles. As previously, the calculations were performed on a remote cluster in parallel overa single node of $48$ cores and took about $1$ hour.
Figure~\ref{fig:argosrmse} shows boxplots of the RMSEs.  The incorporation of domain knowledge significantly improves the performance of the filter for $h=1$, $3$ and $5$ minutes. Figure~\ref{fig:studentmax} shows the same boxplots for the maximum error. Overall, the maximum error is reduced by more than $50\%$ with the penalized particle filters compared to the EKF and penalized EKF for $h= 1$ min. 

Mean values of the maximum error and RMSE over the $20$ trajectories are shown in Table~\ref{tab:argoserrors}.

\begin{table}[ht!]
	\centering
	\begin{tabular}{|ccp{4.5em}p{4.5em}p{4.5em}p{4.5em}p{4.5em}p{4em}|}
		\hline
		$h$ & EKF & Penalized EKF &  Lie-Trotter PF & Penalized Lie-Trotter PF & Strang PF & Penalized Strang PF & Before filter \\
		\hline
		& \multicolumn{7}{c|}{Average max error (km)} \\
		\hline
		$1$ min  & 1.493  & 1.499 & 1.892 & {\bf 0.623} & 2.03 & 0.630 & 6.538\\
		$3$ min & 2.4 & 2.43 & 2.25 & 0.961 & 2.131 & {\bf 0.935} & 5.575\\
		$5$ min & 2.260 & 2.310 & 2.55 & 1.262 & 2.386 & {\bf 1.225} & 3.998\\
		$20$ min & 2.265 & 2.263 &  2.169 & 2.128 & {\bf 2.125} & 5.665 & 2.350\\
		\hline
			& \multicolumn{7}{c|}{Average RMSE (km)} \\
		\hline
		$1$ min  & 0.197 & 0.189  & 0.193 & {\bf 0.156} & 0.189 & {\bf 0.156} & 0.340\\
		$3$ min & 0.267 & 0.257 & 0.259 & {\bf 0.218} & 0.264 & {\bf 0.218} & 0.341\\
		$5$ min & 0.295 & 0.287  & 0.294 & {\bf 0.257}  & 0.295 & 0.258 & 0.340 \\
		$20$ min & 0.348 & 0.347 &  0.408 & 0.353 & 0.396 & {\bf 0.344} & 0.331\\
		\hline
	\end{tabular}
	\caption{Max error and root mean squared error (RMSE) averaged over the $20$ trajectories with Argos X-shaped measurement errors for each value of the time step $h$.  The smallest value in each row is marked in bold. The last column shows the average error in the observations (before filtering) compared to the true positions.
    }
	\label{tab:argoserrors}
\end{table}
These results illustrate particularly well that a precise modelling of the measurement error, together with an accurate consideration of the spatial constraints on the movement, can help to make the most of available data and reconstruct reliable trajectories. Particle filters with domain penalization reduce both RMSE and maximum error compared to Kalman and Extended Kalman filters. However, this gain comes at a computational cost due to the Monte Carlo nature of the particle filter approximation. Overall, the results suggest that EKF-based methods remain effective baseline approaches for Argos-type telemetry, while penalized particle filters offer a way to improve accuracy when computational resources allow and when reliably handling highly non-Gaussian, anisotropic errors is essential.

\section{Application to bowhead whale telemetry data}
\label{sec:bowhead}
 
To illustrate the practical utility of the proposed filtering
framework on real data, we apply it to Argos satellite telemetry
observations of bowhead whales collected
in Hudson Bay and Foxe Basin, Canada  \citep{pomerleau_bowhead_2011}.  Bowhead whales are a species of
conservation concern that inhabit Arctic and sub-Arctic marine waters. Their movement is inherently spatially constrained as their range is bounded by the coastline
geometry of Hudson Bay. 
 
\subsection{Data and pre-processing}
This data were accessed from Movebank (www.movebank.org, study name "Bowhead whale Foxe Basin", Movebank ID 467031755) and are described in \cite{pomerleau_bowhead_2011}. The dataset contains Argos satellite locations for multiple tagged individuals. We focus on individual \texttt{128145} to illustrate our model. This individual is a female tagged in Foxe Basin in June 2013, whose tag transmitted data for 503 days \citep{fortune_age-_2020}.
We retain observations within the bounding box
$[95^{\circ}\text{W},\,70^{\circ}\text{W}]\times
[61^{\circ}\text{N},\,72.5^{\circ}\text{N}]$, remove low-quality fixes
of Argos location class~Z, and discard duplicate timestamps. 
The track comprises $2{,}268$ observations, with a median time step
between successive observations of $1.8$~hours.
Coordinates are projected to the Canada Atlas Lambert system
(EPSG:3978) and expressed in kilometres.
 
The water-domain polygon $\mathcal{D}$ is derived as the complement of
the land surface within the bounding box, using
coastline data from the \texttt{rnaturaleath} package \citep{massicotte_rnaturalearth_2017}.  For simplicity and consistency with the model described in Section~\ref{sec: penalized Langevin}, we retain only one global water polygon, ignoring islands that would represent holes in the polygon.
 
\subsection{Filtering methods}
Movement parameters $(\tau,\,\nu)$ are first estimated by fitting a linear SDE (continuous-time correlated random walk, without the boundary penalty) to the full observation sequence via maximum
likelihood, using the Kalman filter as implemented in the
\texttt{smoothSDE} package \citep{michelot_varying-coefficient_2021}. We obtained the values $\tau=20$ h and $\nu=1.6$ km/h. The angular velocity is set to $0$ since there is visually no clear circular movement in the data.
For the filtering algorithm we set $\nabla H$ to zero, as we are currently not able to estimate it. However, a non-zero potential might be informed by biological knowledge of resting sites or foraging areas. 

We compare two filtering approaches:  
 
\begin{enumerate}
    \item \textbf{Penalised particle filter}: We choose Lie-Trotter
          splitting scheme, spatial penalty strength
          $\lambda = \sqrt{1.8}$, to match the median observation
          interval of $1.8$~hours, and $N = 1000$ particles. For the
Argos measurement error model, we adopt the per-class
degrees-of-freedom and scale parameters of \citet{brost_animal_2015}. When the estimated degree of freedoms were lower than $2$ in \cite{brost_animal_2015}, we instead fixed them to $3$ to ensure that the Gaussian approximations with covariance $\frac{d}{d-2} \Sigma$ are well defined in the particle filter.
    \item \textbf{Kalman filter} :   This corresponds
          to a standard continuous-time correlated random walk
          \citep{johnson_continuous-time_2008} without any spatial penalty ($\lambda \to \infty$). The Argos measurement error is modelled  with a per-class Gaussian distribution with standard deviations based on \citet{brost_animal_2015}.
\end{enumerate}

\subsection{Results}
 
\begin{figure}[ht!]
    \centering

    \includegraphics[width=\textwidth]{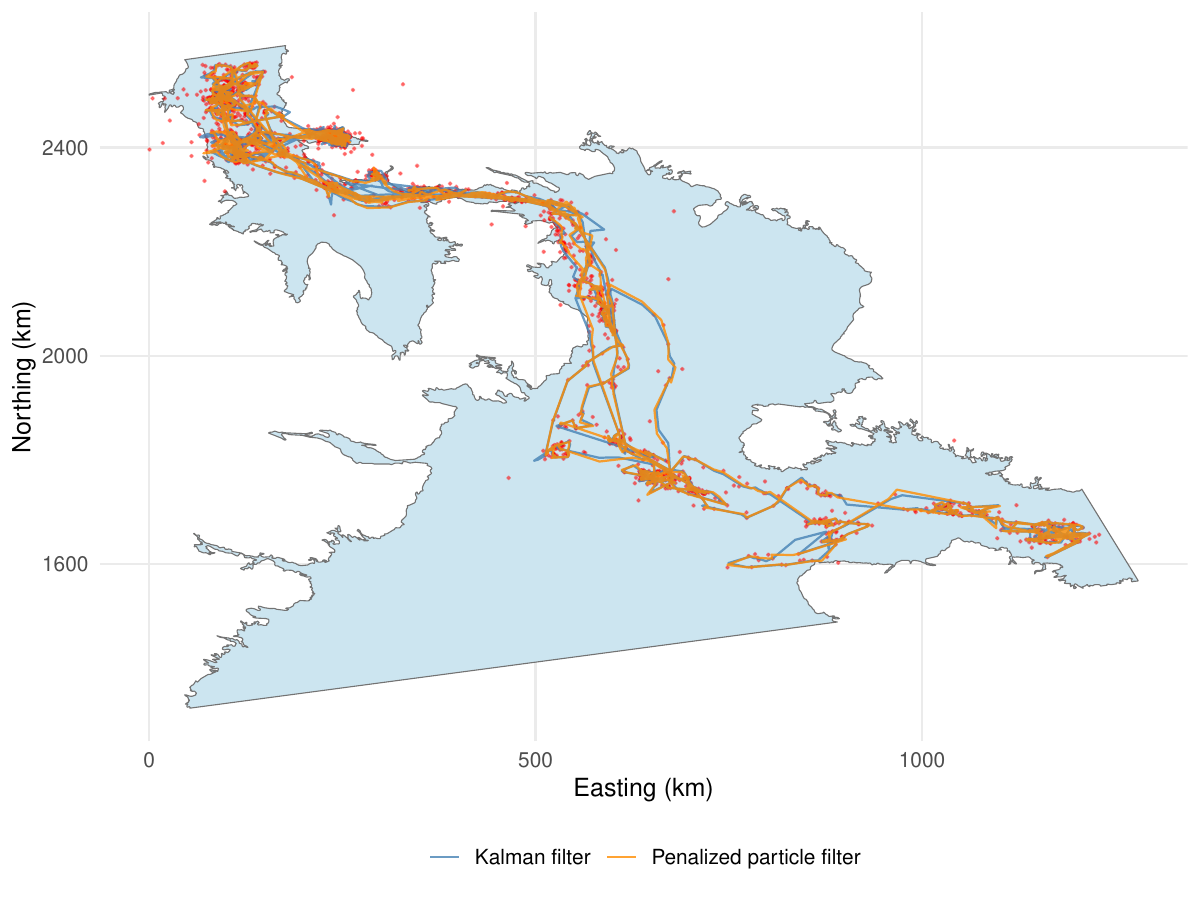}
    \caption{Comparison of filtered tracks for bowhead whale
      ID~128145 in Hudson Bay and Foxe Basin (Canada).  The light blue
      polygon is the water domain $\mathcal{D}$.  Red points are raw
      Argos observations; the orange path is the penalised particle
      filter (Lie-Trotter, $\lambda=\sqrt{1.8}$, $N=1000$); the blue
      path is the Kalman filter (no penalty).}
    \label{fig:bowhead_comparison}
\end{figure}

Figure~\ref{fig:bowhead_comparison} displays the filtered tracks
produced by both methods, overlaid on the water-domain polygon
$\mathcal{D}$ and the raw Argos observations (shown in red). We computed 
the fraction of filtered positions falling outside $\mathcal{D}$, and found that $5.9\%$ ($133$ locations) of Kalman-filtered positions are placed on land,
compared with only $1.6\%$ ($36$ locations) for the penalised particle filter. Hence, the particle filter accommodates better the constraints in the movement. 
We believe this methodology could be broadly applicable to Argos tracking data of whales or other aquatic animals moving in coastal areas.
Regarding computational cost, the particle filtering takes a few minutes to run on a single core, compared to a few seconds for the Kalman filter.
The main cost for the particle filter is the computation of the penalisation term $\beta_\lambda$, which requires to compute a projection onto the polygon boundary. 

\section{Discussion and conclusion}

We introduced a penalized Langevin SDE suitable for modeling the dynamics of animals moving within a constrained domain. This includes potential applications to seals \citep{hanks_reflected_2017}, killer whales \citep{lin_forecasting_2025}, narwhals \citep{delporte_varying_2025} or elk moving in fenced areas \citep{brillinger_simulating_2003}. The effect of landscape boundaries is incorporated through an additional drift term that penalizes trajectories stepping too far outside the domain of interest. This approach was originally proposed by \citet{lions_stochastic_1984} and \citet{liu_discretization_1995} to approximate elliptic reflected SDEs. Compared with \citet{lin_forecasting_2025}, we include the domain constraint directly in the latent dynamics and therefore do not require post-processing of the observed trajectories to ensure land avoidance for aquatic animals.

Simulation of our model relies on splitting the SDE into a non-linear ODE and a linear SDE, both of which can be solved exactly \citep{pilipovic_parameter_2024,pilipovic_strang_2025}. This splitting scheme is straightforward to implement. Based on this approximate solution to the non-linear SDE, we introduced filtering methods for noisy observations of the tracks. We considered the classical Gaussian distribution for the measurement error, as well as more challenging alternatives such as Student's t or mixtures of multivariate Student's t-distributions, which are suitable to model Argos telemetry data errors.

For Gaussian errors, we show that the classical Kalman filter and the Extended Kalman filter perform similarly for high-frequency data. We found that incorporating spatial constraints into the latent movement model within the Kalman filters greatly improves filtering results. Additionally, we proposed a particle filter algorithm to recover true trajectories from observations with heavy-tailed error distributions. We designed Gaussian proposals that approximate the optimal proposal in each case, depending on the splitting scheme used to approximate the SDE. Although they greatly improve filtering accuracy in the case of non-Gaussian measurement error, we emphasize that particle filters come at a considerable computational cost compared with Kalman filtering and may therefore be unsuitable for real-time (online) applications.

The results presented here are application-oriented, and we did not provide theoretical guarantees for the convergence of the Monte Carlo estimation of filtering distributions. Such guarantees can be found in \citet{del_moral_monte-carlo_2001} for the specific case of first-order SDEs solved via the Euler scheme. One of the main requirements for convergence to hold is that the density of the approximation of the solution to the latent SDE converges to the density of the true process. Establishing such a result is particularly difficult, especially when the latent process solves a hypoelliptic SDE (an SDE with degenerate noise), which is constrained in a domain.

A critical parameter in the model is the penalty $\lambda$. A small value of $\lambda$ implies a hard constraint, as the process is immediately pushed back into the domain when leaving it, whereas a large value of $\lambda$ allows the process to leave the boundary for some time and slowly return to the domain of interest. The choice of $\lambda$ should primarily depend on the frequency of the available data. From a theoretical point of view, in first-order reflected SDEs, $\lambda = \sqrt{h}$ is often chosen to minimize the RMSE between the true reflected process and its penalized approximation \citep{pettersson_penalization_1997}. The choice of $\lambda$ may also depend on the level of confidence in the landscape boundaries. If we know with certainty that the animal cannot cross these boundaries, it is sensible to choose a low value of $\lambda$ to ensure that the process remains very close to the ecological domain of interest. Conversely, if there is some uncertainty, a higher value of $\lambda$ can allow the process to leave the domain temporarily.

We implemented the method for domains $\mathcal{D}$ that can be represented as a polygon. However, in practice, the water areas are much more complex than a polygon. Typically, they may be defined as the complement of one or more land polygons, potentially including islands. Hence, the approach would benefit from being extended to such domains. This is mainly an implementation challenge, and would not change the methods described in the paper. 
It is also worth noting that, although our penalization enforces only a soft constraint, hard containment within an original domain $\mathcal{D}$ can be approximated in practice by applying the penalization to a subdomain $\mathcal{D}' \subset \mathcal{D}$, obtained by shrinking $\mathcal{D}$ by a buffer of width $\delta > 0$. Since the process may wander at most a small distance outside $\mathcal{D}'$ before being pushed back inward, an appropriate choice of $\delta$ (relative to $\lambda$) could ensure that the trajectory remains within $\mathcal{D}$. 

We assumed that all parameters were fixed during the filtering, including the movement parameters $\tau$, $\nu$, and $\omega$, as well as the potential surface parameters $x^*$, $\alpha$, and $B$, and the parameters of the measurement error model. In the real data application, we estimated $\tau$, $\nu$ with a standard continuous-time correlated random walk, set $\omega$ to $0$ and the potential to $0$. Even under these simplifying assumptions, the particle filter seems to improve upon the Kalman filter. In practice, the movement parameters and the potential surface need to be estimated from the data, whereas the measurement error parameters are generally known, depending on the type of tracking device attached to the animal. Future work will focus on estimating these parameters from data using Monte Carlo estimates of the filtering distributions obtained from the particle filters.

In the latent dynamics, the potential surface $H$ may depend on environmental covariates that link areas of attraction with feeding habitats. For instance, the presence of whales in a region is strongly influenced by the availability of zooplankton, which is closely associated with patterns of chlorophyll-a productivity \citep{panigada_targeting_2024}.
Such covariate might be included in the latent movement model through resource selection functions \citep{michelot_multiscale_2024}.

Having accurate filtered trajectories is the basis of many analyses of movement tracks. For instance, discrete-time models based on step lengths and turning angles with a latent behaviour switching model require positions at regular time intervals. Such positions may be derived from the filtered positions obtained from our algorithm by linear interpolation or more sophisticated techniques such as diffusion bridges.

\section*{Acknowledgements}
The authors would like to thank the CNRS and IRN Madef for funding A. Delporte, and MIAI - ANR-19-P3IA-0003 for the funding of A Samson.
We also thank the project MATH-AmSud 23-MATH-12, and the UCPH-CNRS 2022 joint PhD programme.

\section*{Data availability statement}
The bowhead whale telemetry data used in this study are publicly available on Movebank (\url{https://www.movebank.org}), under study ID 467031755. The R code to reproduce the analyses is available at \url{github.com/alexandre-delporte/LangevinAnimalTracking}.

\section*{Competing interests}
The authors declare no financial or non-financial competing interests relevant to this work.

\newpage 
\bibliographystyle{plainnat}
\bibliography{biblio}

@article{wensveen_path_2015,
	title = {A path reconstruction method integrating dead-reckoning and position fixes applied to humpback whales},
	volume = {3},
	issn = {2051-3933},
	doi = {10.1186/s40462-015-0061-6},
	language = {en},
	number = {1},
	journal = {Movement Ecology},
	author = {Wensveen, Paul J. and Thomas, Len and Miller, Patrick J. O.},
	year = {2015},
	pages = {31},
}

@article{lions_stochastic_1984,
	title = {Stochastic differential equations with reflecting boundary conditions},
	volume = {37},
	issn = {0010-3640, 1097-0312},
	doi = {10.1002/cpa.3160370408},
	language = {en},
	number = {4},
	urldate = {2024-09-08},
	journal = {Communications on Pure and Applied Mathematics},
	author = {Lions, P. L. and Sznitman, A. S.},
	month = jul,
	year = {1984},
	pages = {511--537},
}

@article{gurarie_correlated_2017,
	title = {Correlated velocity models as a fundamental unit of animal movement: synthesis and applications},
	volume = {5},
	issn = {2051-3933},
	shorttitle = {Correlated velocity models as a fundamental unit of animal movement},
	doi = {10.1186/s40462-017-0103-3},
	language = {en},
	number = {1},
	urldate = {2025-02-18},
	journal = {Movement Ecology},
	author = {Gurarie, Eliezer and Fleming, Christen H. and Fagan, William F. and Laidre, Kristin L. and Hernández-Pliego, Jesús and Ovaskainen, Otso},
	month = dec,
	year = {2017},
	pages = {13},
}

@article{johnson_continuous-time_2008,
	title = {Continuous-time random walk model for animal telemtry data},
	volume = {89},
	issn = {0012-9658, 1939-9170},
	doi = {10.1890/07-1032.1},
	language = {en},
	number = {5},
	urldate = {2025-02-18},
	journal = {Ecology},
	author = {Johnson, Devin S. and London, Joshua M. and Lea, Mary-Anne and Durban, John W.},
	year = {2008},
	pages = {1208--1215},
}

@article{hanks_reflected_2017,
	title = {Reflected {Stochastic} {Differential} {Equation} {Models} for {Constrained} {Animal} {Movement}},
	volume = {22},
	issn = {1085-7117, 1537-2693},
	doi = {10.1007/s13253-017-0291-8},
	language = {en},
	number = {3},
	urldate = {2025-04-04},
	journal = {Journal of Agricultural, Biological and Environmental Statistics},
	author = {Hanks, Ephraim M. and Johnson, Devin S. and Hooten, Mevin B.},
	month = sep,
	year = {2017},
	pages = {353--372},
}

@article{del_moral_monte-carlo_2001,
	title = {The {Monte}-{Carlo} method for filtering with discrete-time observations},
	volume = {120},
	issn = {0178-8051, 1432-2064},
	doi = {10.1007/PL00008786},
	language = {en},
	number = {3},
	urldate = {2025-04-22},
	journal = {Probability Theory and Related Fields},
	author = {Del Moral, Pierre and Jacod, Jean and Protter, Philip},
	month = jul,
	year = {2001},
	pages = {346--368},
}

@article{cholaquidis_level_2020,
	title = {Level sets and drift estimation for reflected {Brownian} motion with drift},
	issn = {1017-0405},
	doi = {10.5705/ss.202018.0211},
	language = {en},
	urldate = {2025-05-01},
	journal = {Statistica Sinica},
	publisher = {Statistica Sinica (Institute of Statistical Science)},
	author = {Cholaquidis, Alejandro and Fraiman, Ricardo and Mordecki, Ernesto and Papalardo, Cecilia},
	year = {2020},
}

@article{brost_animal_2015,
	title = {Animal movement constraints improve resource selection inference in the presence of telemetry error},
	volume = {96},
	issn = {0012-9658, 1939-9170},
	doi = {10.1890/15-0472.1},
	language = {en},
	number = {10},
	urldate = {2025-06-21},
	journal = {Ecology},
	author = {Brost, Brian M. and Hooten, Mevin B. and Hanks, Ephraim M. and Small, Robert J.},
	month = oct,
	year = {2015},
	pages = {2590--2597},
}

@article{liu_discretization_1995,
	title = {Discretization of a class of reflected diffusion processes},
	volume = {38},
	issn = {03784754},
	doi = {10.1016/0378-4754(93)E0072-D},
	language = {en},
	number = {1-3},
	urldate = {2025-06-26},
	journal = {Mathematics and Computers in Simulation},
	author = {Liu, Yingjie},
	month = may,
	year = {1995},
	pages = {103--108},
}

@article{pilipovic_strang_2025,
	title = {Strang {Splitting} for {Parametric} {Inference} in {Second}-order {Stochastic} {Differential} {Equations}},
	volume = {187},
	issn = {03044149},
	doi = {10.1016/j.spa.2025.104650},
	language = {en},
	urldate = {2025-08-05},
	journal = {Stochastic Processes and their Applications},
	author = {Pilipovic, Predrag and Samson, Adeline and Ditlevsen, Susanne},
	month = sep,
	year = {2025},
	pages = {104650},
}

@article{pilipovic_parameter_2024,
	title = {Parameter {Estimation} in {Nonlinear} {Multivariate} {Stochastic} {Differential} {Equations} {Based} on {Splitting} {Schemes}},
	volume = {52},
	issn = {0090-5364},
	doi = {10.1214/24-AOS2371},
	language = {en},
	number = {2},
	urldate = {2025-08-05},
	journal = {The Annals of Statistics},
	author = {Pilipovic, Predrag and Samson, Adeline and Ditlevsen, Susanne},
	month = apr,
	year = {2024},
}

@article{patterson_using_2010,
	title = {Using {GPS} data to evaluate the accuracy of state–space methods for correction of {Argos} satellite telemetry error},
	volume = {91},
	issn = {0012-9658, 1939-9170},
	doi = {10.1890/08-1480.1},
	language = {en},
	number = {1},
	urldate = {2025-08-29},
	journal = {Ecology},
	author = {Patterson, Toby A. and McConnell, Bernie J. and Fedak, Mike A. and Bravington, Mark V. and Hindell, Mark A.},
	month = jan,
	year = {2010},
	pages = {273--285},
}

@incollection{doucet_tutorial_2011,
	title = {A tutorial on particle filtering and smoothing: fifteen years later},
	isbn = {978-0-19-953290-2},
	booktitle = {The {Oxford} {Handbook} of nonlinear filtering},
	publisher = {Oxford University Press},
	author = {Doucet, Arnaud and Johansen, Adam},
	year = {2011},
	pages = {656--705},
}

@incollection{brillinger_simulating_2003,
	address = {Beachwood, OH},
	title = {Simulating constrained animal motion using stochastic differential equations},
	isbn = {978-0-940600-55-3},
	doi = {10.1214/lnms/1215091656},
	language = {en},
	urldate = {2025-10-07},
	booktitle = {Institute of {Mathematical} {Statistics} {Lecture} {Notes} - {Monograph} {Series}},
	publisher = {Institute of Mathematical Statistics},
	author = {Brillinger, David R.},
	year = {2003},
	pages = {35--48},
}

@article{lin_forecasting_2025,
	title = {Forecasting trajectories of {Southern} {Resident} killer whales with stochastic movement models incorporating direction modification},
	volume = {509},
	issn = {03043800},
	doi = {10.1016/j.ecolmodel.2025.111254},
	language = {en},
	urldate = {2025-10-15},
	journal = {Ecological Modelling},
	author = {Lin, Teng-Wei and Dowd, Michael and Joy, Ruth},
	month = oct,
	year = {2025},
	pages = {111254},
}

@article{jonsen_continuous-time_2020,
	title = {A continuous-time state-space model for rapid quality control of argos locations from animal-borne tags},
	volume = {8},
	issn = {2051-3933},
	doi = {10.1186/s40462-020-00217-7},
	language = {en},
	number = {1},
	urldate = {2025-11-06},
	journal = {Movement Ecology},
	author = {Jonsen, Ian D. and Patterson, Toby A. and Costa, Daniel P. and Doherty, Philip D. and Godley, Brendan J. and Grecian, W. James and Guinet, Christophe and Hoenner, Xavier and Kienle, Sarah S. and Robinson, Patrick W. and Votier, Stephen C. and Whiting, Scott and Witt, Matthew J. and Hindell, Mark A. and Harcourt, Robert G. and McMahon, Clive R.},
	month = dec,
	year = {2020},
	pages = {31},
}

@article{hoenner_enhancing_2012,
	title = {Enhancing the {Use} of {Argos} {Satellite} {Data} for {Home} {Range} and {Long} {Distance} {Migration} {Studies} of {Marine} {Animals}},
	volume = {7},
	issn = {1932-6203},
	doi = {10.1371/journal.pone.0040713},
	language = {en},
	number = {7},
	urldate = {2025-11-06},
	journal = {PLoS ONE},
	author = {Hoenner, Xavier and Whiting, Scott D. and Hindell, Mark A. and McMahon, Clive R.},
	editor = {Hays, Graeme Clive},
	month = jul,
	year = {2012},
	pages = {e40713},
}

@article{hays_high_2021,
	title = {High accuracy tracking reveals how small conservation areas can protect marine megafauna},
	volume = {31},
	issn = {1051-0761, 1939-5582},
	doi = {10.1002/eap.2418},
	language = {en},
	number = {7},
	urldate = {2025-11-06},
	journal = {Ecological Applications},
	author = {Hays, Graeme C. and Mortimer, Jeanne A. and Rattray, Alex and Shimada, Takahiro and Esteban, Nicole},
	month = oct,
	year = {2021},
	pages = {e02418},
}

@article{panigada_targeting_2024,
	title = {Targeting fin whale conservation in the {North}-{Western} {Mediterranean} {Sea}: insights on movements and behaviour from biologging and habitat modelling},
	volume = {11},
	issn = {2054-5703},
	shorttitle = {Targeting fin whale conservation in the {North}-{Western} {Mediterranean} {Sea}},
	doi = {10.1098/rsos.231783},
	language = {en},
	number = {3},
	urldate = {2025-11-13},
	journal = {Royal Society Open Science},
	author = {Panigada, Viola and Bodey, Thomas W. and Friedlaender, Ari and Druon, Jean-Noël and Huckstädt, Luis A. and Pierantonio, Nino and Degollada, Eduard and Tort, Beatriu and Panigada, Simone},
	month = mar,
	year = {2024},
	pages = {231783},
}

@article{pettersson_penalization_1997,
	title = {Penalization {Schemes} for {Reflecting} {Stochastic} {Differential} {Equations}},
	volume = {3},
	issn = {13507265},
	doi = {10.2307/3318456},
	number = {4},
	urldate = {2025-11-18},
	journal = {Bernoulli},
	author = {Pettersson, Roger},
	month = dec,
	year = {1997},
	pages = {403},
}

@article{preisler_modeling_2004,
	title = {Modeling animal movements using stochastic differential equations},
	volume = {15},
	issn = {1180-4009, 1099-095X},
	doi = {10.1002/env.636},
	language = {en},
	number = {7},
	urldate = {2025-11-18},
	journal = {Environmetrics},
	author = {Preisler, Haiganoush K. and Ager, Alan A. and Johnson, Bruce K. and Kie, John G.},
	month = nov,
	year = {2004},
	pages = {643--657},
}

@misc{michelot_multiscale_2024,
	title = {Multiscale modelling of animal movement with persistent dynamics},
	copyright = {Creative Commons Attribution 4.0 International},
	doi = {10.48550/ARXIV.2406.15195},
	urldate = {2025-11-18},
	publisher = {arXiv},
	author = {Michelot, Théo and Hanks, Ephraim M.},
	year = {2024},
	note = {Version Number: 2},
}

@article{michelot_varying-coefficient_2021,
	title = {Varying-{Coefficient} {Stochastic} {Differential} {Equations} with {Applications} in {Ecology}},
	volume = {26},
	issn = {1085-7117, 1537-2693},
	doi = {10.1007/s13253-021-00450-6},
	language = {en},
	number = {3},
	urldate = {2025-11-18},
	journal = {Journal of Agricultural, Biological and Environmental Statistics},
	author = {Michelot, Théo and Glennie, Richard and Harris, Catriona and Thomas, Len},
	month = sep,
	year = {2021},
	pages = {446--463},
}

@article{russell_spatially_2018,
	title = {A spatially varying stochastic differential equation model for animal movement},
	volume = {12},
	issn = {1932-6157},
	doi = {10.1214/17-AOAS1113},
	language = {en},
	number = {2},
	urldate = {2026-02-28},
	journal = {The Annals of Applied Statistics},
	author = {Russell, James C. and Hanks, Ephraim M. and Haran, Murali and Hughes, David},
	month = jun,
	year = {2018},
}

@article{albertsen_generalizing_2019,
	title = {Generalizing the first-difference correlated random walk for marine animal movement data},
	volume = {9},
	issn = {2045-2322},
	doi = {10.1038/s41598-019-40405-z},
	language = {en},
	number = {1},
	urldate = {2026-03-17},
	journal = {Scientific Reports},
	author = {Albertsen, Christoffer Moesgaard},
	month = mar,
	year = {2019},
	pages = {4017},
}

@article{pomerleau_bowhead_2011,
	title = {Bowhead whale {Balaena} mysticetus diving and movement patterns in the eastern {Canadian} {Arctic}: implications for foraging ecology},
	volume = {15},
	issn = {1863-5407, 1613-4796},
	shorttitle = {Bowhead whale {Balaena} mysticetus diving and movement patterns in the eastern {Canadian} {Arctic}},
	doi = {10.3354/esr00373},
	language = {en},
	number = {2},
	urldate = {2026-03-24},
	journal = {Endangered Species Research},
	author = {Pomerleau, C and Patterson, Ta and Luque, S and Lesage, V and Heide-Jørgensen, Mp and Dueck, Ll and Ferguson, Sh},
	month = nov,
	year = {2011},
	pages = {167--177},
}

@misc{massicotte_rnaturalearth_2017,
	title = {rnaturalearth: {World} {Map} {Data} from {Natural} {Earth}},
	shorttitle = {rnaturalearth},
	doi = {10.32614/CRAN.package.rnaturalearth},
	language = {en},
	urldate = {2026-03-25},
	author = {Massicotte, Philippe and South, Andy},
	month = mar,
	year = {2017},
	note = {Institution: Comprehensive R Archive Network
Pages: 1.2.0},
}

@article{fortune_age-_2020,
	title = {Age- and sex-specific movement, behaviour and habitat-use patterns of bowhead whales ({Balaena} mysticetus) in the {Eastern} {Canadian} {Arctic}},
	volume = {43},
	issn = {0722-4060, 1432-2056},
	doi = {10.1007/s00300-020-02739-7},
	language = {en},
	number = {11},
	urldate = {2026-03-26},
	journal = {Polar Biology},
	author = {Fortune, Sarah M. E. and Young, Brent G. and Ferguson, Steven H.},
	month = nov,
	year = {2020},
	pages = {1725--1744},
}

@article{gloaguen_stochastic_2018,
	title = {Stochastic {Differential} {Equation} {Based} on a {Multimodal} {Potential} to {Model} {Movement} {Data} in {Ecology}},
	volume = {67},
	issn = {0035-9254, 1467-9876},
	doi = {10.1111/rssc.12251},
	language = {en},
	number = {3},
	urldate = {2026-04-01},
	journal = {Journal of the Royal Statistical Society Series C: Applied Statistics},
	author = {Gloaguen, Pierre and Etienne, Marie-Pierre and Le Corff, Sylvain},
	month = apr,
	year = {2018},
	pages = {599--619},
}

@article{kalman_general_1960,
	title = {On the general theory of control systems},
	volume = {1},
	issn = {14746670},
	doi = {10.1016/S1474-6670(17)70094-8},
	language = {en},
	number = {1},
	urldate = {2026-04-06},
	journal = {IFAC Proceedings Volumes},
	author = {Kalman, R.E.},
	month = aug,
	year = {1960},
	pages = {491--502},
}

@article{mclachlan_splitting_2002,
	title = {Splitting methods},
	volume = {11},
	copyright = {https://www.cambridge.org/core/terms},
	issn = {0962-4929, 1474-0508},
	doi = {10.1017/S0962492902000053},
	language = {en},
	urldate = {2026-04-08},
	journal = {Acta Numerica},
	author = {McLachlan, Robert I. and Quispel, G. Reinout W.},
	month = jan,
	year = {2002},
	pages = {341--434},
}

@misc{maechler_expm_2010,
	title = {expm: {Matrix} {Exponential}, {Log}, 'etc'},
	shorttitle = {expm},
	url = {https://CRAN.R-project.org/package=expm},
	doi = {10.32614/CRAN.package.expm},
	language = {en},
	urldate = {2026-04-08},
	author = {Maechler, Martin and Dutang, Christophe and Goulet, Vincent},
	month = jul,
	year = {2010},
	note = {Institution: Comprehensive R Archive Network
Pages: 1.0-0},
}

@article{michelot_langevin_2019,
	title = {The {Langevin} diffusion as a continuous‐time model of animal movement and habitat selection},
	volume = {10},
	issn = {2041-210X, 2041-210X},
	doi = {10.1111/2041-210X.13275},
	language = {en},
	number = {11},
	urldate = {2026-04-08},
	journal = {Methods in Ecology and Evolution},
	author = {Michelot, Théo and Gloaguen, Pierre and Blackwell, Paul G. and Étienne, Marie‐Pierre},
	editor = {Ergon, Torbjørn},
	month = nov,
	year = {2019},
	pages = {1894--1907},
}

@book{hooten_animal_2017,
	address = {Boca Raton : CRC Press, 2017.},
	edition = {1},
	title = {Animal {Movement}: {Statistical} {Models} for {Telemetry} {Data}},
	isbn = {978-1-315-11774-4},
	shorttitle = {Animal {Movement}},
	doi = {10.1201/9781315117744},
	language = {en},
	urldate = {2026-04-10},
	publisher = {CRC Press},
	author = {Hooten, Mevin B. and Johnson, Devin S. and McClintock, Brett T. and Morales, Juan M.},
	month = mar,
	year = {2017},
}

@article{delporte_varying_2025,
	title = {Varying coefficients correlated velocity models in complex landscapes with boundaries applied to narwhal responses to noise exposure},
	volume = {19},
	issn = {1932-6157},
	doi = {10.1214/25-AOAS2070},
	number = {4},
	urldate = {2026-04-21},
	journal = {The Annals of Applied Statistics},
	author = {Delporte, Alexandre and Ditlevsen, Susanne and Samson, Adeline},
	month = dec,
	year = {2025},
}

@article{eddelbuettel_rcpp_2011,
	title = {\textbf{{Rcpp}} : {Seamless} \textit{{R}} and \textit{{C}++} {Integration}},
	volume = {40},
	issn = {1548-7660},
	doi = {10.18637/jss.v040.i08},
	language = {en},
	number = {8},
	urldate = {2026-05-04},
	journal = {Journal of Statistical Software},
	author = {Eddelbuettel, Dirk and François, Romain},
	year = {2011},
}

\newpage
\appendix

\section{Derivation of the gradient and Hessian of the potential}\label{App:A}
We derive the gradient and Hessian matrix of the potential \[H(x)=-\sum_{j=1}^J H_j(x) \mbox{ with } H_j(x)=\alpha_j \exp(-(x-x^*_j)^\top B_j(x-x^*_j)).\]
We use the following formulas for matrix and vector differentiation.
For any $\phi : \R \rightarrow \R$, $a :\R^d \rightarrow \R$, $v :\R^d \rightarrow \R^d$, $B \in M_{d,d}(\R)$,
\begin{equation}\nabla\phi(a(x))=\phi'(a(x))\nabla a(x) 
\label{eq: chain rule}
\end{equation}
\begin{equation} \quad \nabla(x^\top Bx)=(B+B^\top) x 
\label{eq: gradient quadratic form}
\end{equation}
\begin{equation} \quad D(a(x)v(x))= v(x)\nabla a(x)^\top+a(x) Dv(x) \label{eq: product rule} \end{equation}
where $D$ denotes the Jacobian matrix.

By \eqref{eq: chain rule} and \eqref{eq: gradient quadratic form} we get
\begin{align*}
	\nabla H_j(x)&=-\alpha_j(B_j+B_j^\top) (x-x_j^*) \exp(-(x-x_j^*)^\top B_j(x-x_j^*)) \\
	&= -2\alpha_jB_j (x-x_j^*) \exp(-(x-x_j^*)^\top B_j(x-x_j^*))
\end{align*}
where we used that $B_j$ is symmetric.
Then, by \eqref{eq: product rule},
\begin{align*}
	D^2H_j(x)& =D\nabla H_j(x)=-2\alpha_j\Bigg( B_j(x-x_j^*) \times \left(-2B_j(x-x_j^*) \exp(-(x-x_j^*)^\top B_j(x-x_j^*))\right)^\top \\
	& \quad \quad + \exp(-(x-x_j^*)^\top B_j(x-x_j^*)) \times B_j\Bigg) \\
	&=-2\alpha_j\Bigg( -2B_j(x-x_j^*)(x-x_j^*)^\top B_j^\top \exp(-(x-x_j^*)^\top B_j(x-x_j^*)) \\
	& \quad \quad + \exp(-(x-x_j^*)^\top B_j(x-x_j^*)) \times B_j\Bigg) \\
	&=-2\alpha_j \exp(-(x-x_j^*)^\top B_j(x-x_j^*))(B_j-2B_j (x-x_j^*)(x-x_j^*)^\top B_j) \\
	&= -2\alpha_j \exp(-(x-x_j^*)^\top B_j(x-x_j^*))B_j(I_2-2 (x-x_j^*)(x-x_j^*)^\top B_j). 
\end{align*}

\section{Full calculation for SDE splitting}\label{App:B}
Let $l \in \{1,\cdots,J\}$.
We want to split the following non-linear SDE:
\begin{equation}
	\begin{cases}
		dX(t)=V(t) dt \\
		dV(t)=-AV(t)+\nabla H_l(X(t))dt-\nabla H_{-l}(X(t)) dt+\sigma dW(t)-\beta_\lambda(X(t)) dt
	\end{cases}
\label{eq: penalized Langevin for splitting}
\end{equation}
where $H_{-l}(x)=-\sum_{j=1, j \neq l}^J H_j(x)$.
We define \begin{align*}F(x,v)&=-Av+\nabla H_l(x)dt-\beta_\lambda(x) \\
&= -Av-2\alpha_lB_l(x-x_l^*)e_l(x)-\beta_\lambda(x)\end{align*}
where $e_l(x)=\exp(-(x-x_l^*)^\top B_l(x-x_l^*))$.

\eqref{eq: penalized Langevin for splitting} can be rewritten as 
\begin{equation}
    dU(t)=\tilde{F}(U(t))dt-\begin{pmatrix} 0_2 \\ \nabla H_{-l}(X(t)) \end{pmatrix} dt +\tilde{\Sigma} dW(t)
\end{equation}
with $\tilde{F}(x,v)=\begin{pmatrix} v & F(x,v) \end{pmatrix}^\top$

Hence, we can decompose the equation as 
\[dU(t)=\tilde{A}(U(t)-u^*)dt +\tilde{N}(x,v)+\tilde{\Sigma} dW(t)\]
with $\tilde{A}=\begin{pmatrix}  0_2 & I_2 \\ A_x & A_v \end{pmatrix}$, $ \tilde{N}(x,v)=\begin{pmatrix} 0_2 \\ N(x,v)-\nabla H_{-l}(X(t))\end{pmatrix}$, $u^*=\begin{pmatrix} x^* \\ 0_2\end{pmatrix} $ and $A_x$, $A_v$, $x^*$, $N(x,v)$ such that $$F(x,v)=A_x(x-x^*)+A_vv+N(x,v)$$

It is clear that $u_l^*=\begin{pmatrix}  x_l^* & 0_{1,2} \end{pmatrix}$ is a zero of $\tilde{F}$.
We split the drift as follows : $ \tilde{F}(x,v)=\tilde{A}(u-u_l^*) +\tilde{N}(x,v)$ with 
\[\tilde{A}=D_u  \tilde{F}(u_l^*)= \begin{pmatrix} 0_2 & I_2 \\ -2\alpha_lB_l & -A\end{pmatrix},\quad \tilde{N}(x,v)=\begin{pmatrix} 0_{2,1} \\ N(x,v)\end{pmatrix}\]
and 
\begin{align*}
    N(x,v)&=F(x,v)-A_x(x-x^*_l)-A_vv\\
    &=-Av-2\alpha_lB_l(x-x_l^*)e_l(x) -\beta_\lambda(x)+2\alpha_lB_l(x-x_l^*)+Av \\
    &= -2\alpha_lB_l(x-x_l^*)(e_l(x)-1)- \beta_{\lambda}(x)
\end{align*}

Hence we obtain equation \eqref{eq: penalized Langevin rewritten}.

\section{Covariance matrix for the OU process}\label{App:C}
The covariance matrix of the OU process in the splitting is:
$$\tilde{Q}(h)=\int_0^h e^{\tilde{A}u} \Gamma   e^{-\tilde{A}^\top u} du .$$
Following \cite{albertsen_generalizing_2019}, we vectorize the covariance matrix,
\begin{align*}
	\mathrm{vec}(\tilde{Q}(h))&=\int_0^h e^{\tilde{A}u}\otimes e^{\tilde{A}u} \mathrm{vec}(\Gamma) du \\
		&=\int_0^h e^{(\tilde{A} \oplus \tilde{A})u} du  \, \mathrm{vec}(\Gamma) \\
		&= (\tilde{A} \oplus \tilde{A})^{-1} (e^{(\tilde{A} \oplus \tilde{A}) h}-I_{16}) \,\mathrm{vec}(\Gamma) \\
        &=(\tilde{A} \oplus \tilde{A})^{-1} e^{(\tilde{A} \oplus \tilde{A}) h}\mathrm{vec}(\Gamma)-(\tilde{A} \oplus \tilde{A})^{-1} \,\mathrm{vec}(\Gamma) \\
        &=e^{(\tilde{A} \oplus \tilde{A}) h} (\tilde{A} \oplus \tilde{A})^{-1} \mathrm{vec}(\Gamma)-(\tilde{A} \oplus \tilde{A})^{-1} \,\mathrm{vec}(\Gamma)
\end{align*}
where we used the following properties of the vectorization operator and Kronecker sum and product: for any matrices $A$,$B$ and $C$, $$\mathrm{vec}(ABC)=(C^\top \otimes A) \mathrm{vec}(B)$$
$$e^{M} \otimes e^{N}=e^{M \oplus N}.$$
Let $C$ be the matrix such that $\mathrm{vec}(C)=(\tilde{A} \oplus \tilde{A})^{-1} \mathrm{vec}(\Gamma)$. Then 
\begin{align*}
    \mathrm{vec}(\tilde{Q}(h))=e^{(\tilde{A} \oplus \tilde{A}) h} \, \mathrm{vec}(C)-\mathrm{vec}(C) = e^{\tilde{A}h}  \otimes e^{\tilde{A} h} \, \mathrm{vec}(C)- \mathrm{vec}(C) = \mathrm{vec}(e^{\tilde{A}h} C e^{\tilde{A}^\top h}-C)
\end{align*}
Hence $\tilde{Q}(h)=e^{\tilde{A}h} C e^{\tilde{A}^\top h}-C$.
 
\end{document}